\documentclass[pdflatex,sn-mathphys-ay]{sn-jnl}% Math and Physical Sciences Author-Year Reference Style

\usepackage{graphicx}%
\usepackage{multirow}%
\usepackage{amsmath,amssymb,amsfonts}%
\usepackage{mathrsfs}%
\usepackage{xcolor}%
\usepackage{textcomp}%
\usepackage{manyfoot}%
\usepackage{booktabs}%
\usepackage{makecell}%
\usepackage{algorithm}%
\usepackage{algorithmic}%
\usepackage{listings}%

\definecolor{codegreen}{rgb}{0.25,0.50,0.35}
\definecolor{codegray}{rgb}{0.5,0.5,0.5}
\definecolor{codepurple}{rgb}{0.58,0,0.82}
\definecolor{backcolour}{rgb}{0.97,0.97,0.97}

\lstdefinestyle{elegantpython}{
    language=Python,
    backgroundcolor=\color{backcolour},
    commentstyle=\color{codegreen}\itshape,
    keywordstyle=\color{blue}\bfseries,
    numberstyle=\tiny\color{codegray},
    stringstyle=\color{codepurple},
    basicstyle=\ttfamily\footnotesize,
    breakatwhitespace=false,
    breaklines=true,
    captionpos=b,
    keepspaces=true,
    numbers=left,
    numbersep=8pt,
    showspaces=false,
    showstringspaces=false,
    showtabs=false,
    tabsize=4,
    frame=single,
    rulecolor=\color{black!20},
    framesep=3pt,
    xleftmargin=12pt,
}
\begin{document}

\title[A 99-Line Homogenization Code for Lattice-skin Plate Structures]{A 99-Line Homogenization Code for Lattice-skin Plate Structures}

\author[1]{\fnm{Zhongkai} \sur{Ji}}
\author*[2]{\fnm{Dawei} \sur{Li}}\email{ldw@njust.edu.cn}
\author*[1]{\fnm{Yong} \sur{Zhao}}\email{yongzhao@sjtu.edu.cn}
\author[2]{\fnm{Wenhe} \sur{Liao}}

\affil[1]{\orgname{Shanghai Key Laboratory of Digital Manufacture for Thin-Walled Structures, Shanghai Jiao Tong University}, \orgaddress{\city{Shanghai}, \postcode{200240}, \country{PR China}}}
\affil[2]{\orgname{School of Mechanical Engineering, Nanjing University of Science and Technology}, \orgaddress{\city{Nanjing}, \postcode{210094}, \country{PR China}}}

\abstract{
Recent years have seen growing application potential for Lattice-skin Plate Structures in advanced manufacturing fields such as aerospace and automotive engineering. For multiscale performance evaluation of such structures, conventional homogenization methods for lattice-filled volume structures are often used for equivalent analysis. However, in finite-thickness Lattice-skin Plate Structures, periodic boundary conditions imposed along the three orthogonal directions of the representative cell cannot adequately capture the boundary effect of the free surfaces in the thickness direction, which introduces bias into the prediction of effective properties. To reduce this bias, this study develops and open-sources a homogenization method for Lattice-skin Plate Structures, forming an open-source computational framework for this class of structures. Representative numerical examples show that the framework can stably extract effective plate/shell stiffness matrices and can be extended to predict multiphase material properties and analyze steady-state heat conduction. The tool provides an open and reusable analysis foundation for the high-fidelity design of multifunctional lightweight structures.
}


\maketitle
\section{Introduction}
As a new class of lightweight structures, Lattice-skin Plate Structures show application potential in advanced manufacturing fields such as aerospace\citep{khanSystematicReviewDesign2024, sunRigidflexibleInterlockedMetastructures2025} and automotive engineering\citep{wangBioinspiredVertexoffsetLattice2025}. For performance evaluation of such multiscale structures, the homogenization method for lattice-filled volume structures (LVS-H) is commonly used for multiscale dimensional reduction analysis\citep{bisharaStateoftheArtReviewMachine2023}. As skin-enabled structures continue to become thinner, LVS-H may introduce free-surface truncation errors when applied to finite-thickness structures, which leads to bias in the extracted effective mechanical properties\citep{yiFEMFormulationHomogenization2015}. This is because LVS-H imposes periodic boundary conditions along the three orthogonal directions of the representative volume element (RVE). The mathematical assumption of periodic continuation in the volume does not adequately reflect the free-surface mechanical state through the thickness of a thin plate, and may therefore overestimate the tensile and bending stiffness of the structure\citep{caiNovelNumericalImplementation2014}.

To reduce the bias in effective-property prediction for thin plates, the homogenization method for Lattice-skin Plate Structures (LPS-H) has been developed. In the analysis of composites and thin-walled structures, LPS-H preserves free boundary conditions in the out-of-plane normal direction and can characterize plate/shell responses such as extension-bending and extension-twist coupling\citep{caiNovelNumericalImplementation2014, qiaoNumericalImplementationVariational2021, zhouEfficientMethodEstimate2025}. However, for finite-thickness Lattice-skin Plate Structures with complex topological porosity, LVS-H is still commonly used at present.

\begin{figure*}[htbp]
  \centering
  \includegraphics[width=\textwidth]{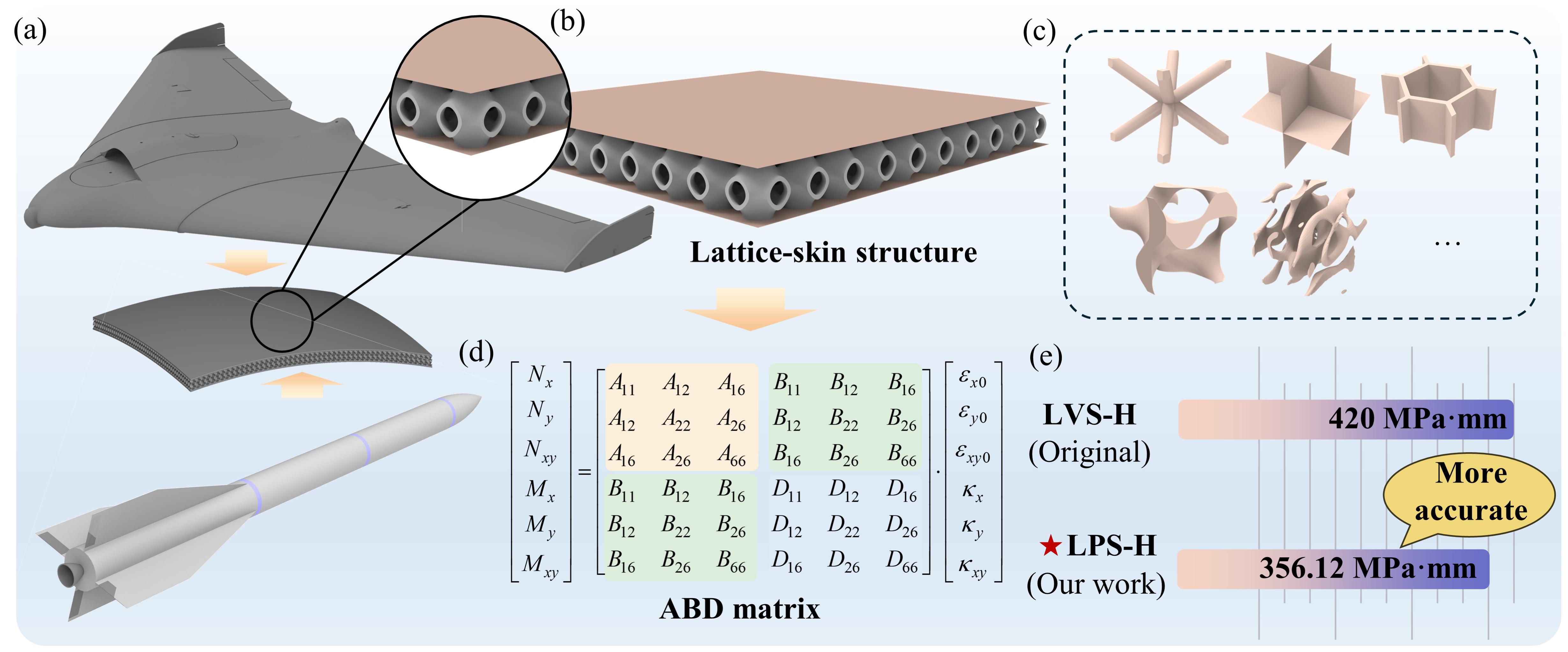}
  \caption{Open-source LPS-H analysis framework for Lattice-skin Plate Structures. (a,b) Typical applications of Lattice-skin Plate Structures in aerospace fields such as aircraft wings and missile shells. (c) Lightweight Lattice-skin Plate Structures with multiple configurations. (d) Definition of the plate/shell ABD stiffness matrix. (e) Comparison of LPS-H and LVS-H in the prediction of macroscopic effective stiffness.}
  \label{fig:fig_overview}
\end{figure*}

Meanwhile, in computational mechanics and structural optimization, open-source programs and educational papers play an important supporting role in the dissemination of technical methods and the development of new algorithms. For example, volume homogenization methods for analyzing the effective elastic tensors of two-dimensional (2D)\citep{andreassenHowDetermineComposite2014} and three-dimensional (3D) structures\citep{dong149LineHomogenization2019}, together with foundational codes for structural topology optimization\citep{sigmund99LineTopology2001,zhouComplementaryLectureNotes2021, woldseth808LinePhasorbased2024, dingEasytouseUnivariateMappingbased2025}, have been used in many follow-up studies\citep{hanMultimaterialTopologyOptimization2022, gaoRationalDesignsMechanical2023, bayatHolisticComputationalDesign2023, zhengDeepLearningMechanical2023} and have provided a basis for secondary development of related algorithms. Motivated by these efforts, the present work provides a compact and efficient open-source framework implementing LPS-H, as shown in \textbf{Figure \ref{fig:fig_overview}}, to supplement the open-source tools available for the mechanical analysis of Lattice-skin Plate Structures. This framework introduces loop-free multidimensional tensor mapping and graphics processing unit (GPU) heterogeneous acceleration, and serves as a core computational tool for the high-fidelity mechanical evaluation and design of multifunctional lightweight Lattice-skin Plate Structures.

\section{Theoretical Frameworks for LPS-H}
This section first briefly reviews the macroscopic characterization of plate/shell mechanical properties and then compares the theoretical differences between LVS-H and LPS-H in predicting the effective stiffness of complex Lattice-skin Plate Structures, thereby establishing the mathematical basis for the open-source implementation presented later.

\subsection{Macroscopic plate mechanics and the ABD stiffness matrix}

\begin{figure*}[htbp]
  \centering
  \includegraphics[width=\textwidth]{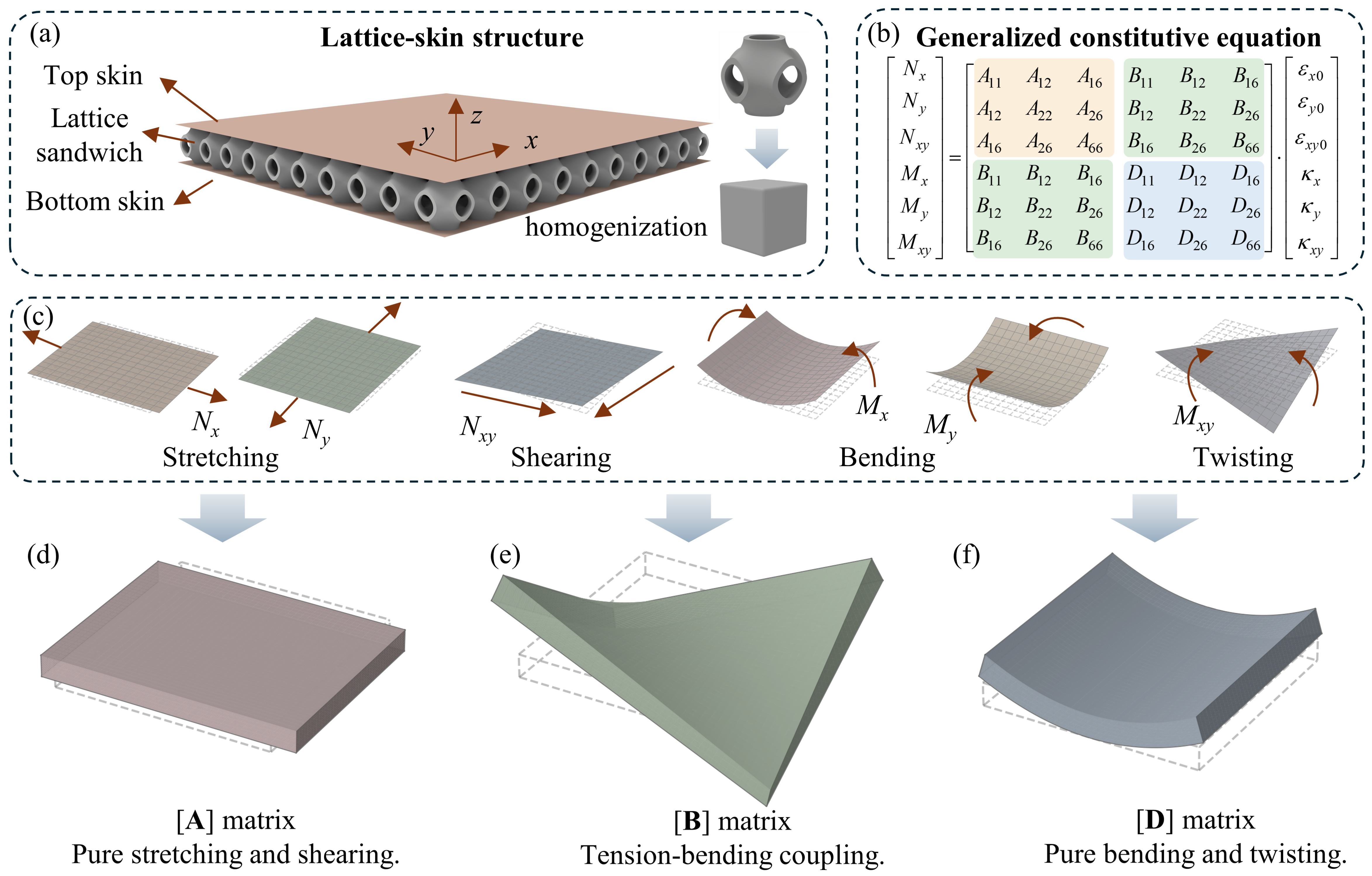}
  \caption{Equivalent plate/shell mechanical characterization of Lattice-skin Plate Structures and schematic illustration of the ABD stiffness matrix. (a) Dimensional reduction from a Lattice-skin Plate Structure to an equivalent homogeneous plate/shell. (b) Composition and definition of the ABD stiffness matrix. (c) Six basic loading forms used to extract the effective stiffness components. (d-f) Physical meanings of the stiffness components in the $\mathbf{A}$, $\mathbf{B}$, and $\mathbf{D}$ matrices.}
  \label{fig:fig1}
\end{figure*}

In solid mechanics analysis, full-scale modeling of Lattice-skin Plate Structures with complex internal porosity using a high-fidelity 3D finite element model usually leads to high computational cost\citep{zhangOptimizedEasytouseOpensource2023, yangGuidedDiffusionFast2024}. Therefore, replacing the detailed structure with an equivalent macroscopic homogeneous plate/shell is a common dimensional-reduction strategy in engineering practice, as shown in \textbf{Figure \ref{fig:fig1}a}. To achieve this cross-scale mechanical equivalence, the basic relation between 3D continuum mechanics and macroscopic plate/shell mechanics must be clarified.

Within the framework of Classical Laminate Plate Theory (CLPT) or First-Order Shear Deformation Theory (FSDT), the macroscopic kinematics of a plate/shell are mainly governed by deformation of the mid-surface. Under the plane-section assumption, the in-plane strain component $\boldsymbol{\varepsilon}(z)$ at an arbitrary point located at the normal coordinate $z$ from the geometric mid-surface can be expressed as the linear superposition of mid-surface membrane deformation and shear, together with bending and twisting deformation:

\begin{equation}
    \boldsymbol{\varepsilon}(z) = \boldsymbol{\varepsilon}^0 + z \boldsymbol{\kappa}
    \label{eq:plate_kinematics}
\end{equation}

Here, $\boldsymbol{\varepsilon}^0 = [\varepsilon_{11}^0, \varepsilon_{22}^0, \gamma_{12}^0]^T$ denotes the mid-surface membrane strain and describes the in-plane stretching and shear of the plate/shell. Likewise, $\boldsymbol{\kappa} = [\kappa_{11}, \kappa_{22}, \kappa_{12}]^T$ denotes the mid-surface curvature and describes the out-of-plane bending and twisting of the plate/shell.

For a thin plate under plane stress, the linear elastic constitutive relation between the local microscopic in-plane stress $\boldsymbol{\sigma}_p = [\sigma_{11}, \sigma_{22}, \sigma_{12}]^T$ and the local strain $\boldsymbol{\varepsilon}(z)$ is governed by the reduced plane-stress stiffness matrix $\mathbf{Q}(z)$:

\begin{equation}
    \boldsymbol{\sigma}_p(z) = \mathbf{Q}(z) \boldsymbol{\varepsilon}(z) = \mathbf{Q}(z) (\boldsymbol{\varepsilon}^0 + z \boldsymbol{\kappa})
    \label{eq:local_stress}
\end{equation}

At the macroscopic level, the internal resultants of the plate/shell are obtained by integrating the local in-plane stress $\boldsymbol{\sigma}_p$ through the physical thickness $h$, from $z=-h/2$ to $z=h/2$, which yields the equivalent membrane force $\mathbf{N}$ and bending moment $\mathbf{M}$:

\begin{equation}
    \mathbf{N} = \int_{-h/2}^{h/2} \boldsymbol{\sigma}_p(z) dz, \quad \mathbf{M} = \int_{-h/2}^{h/2} \boldsymbol{\sigma}_p(z) z dz
    \label{eq:stress_resultants}
\end{equation}

Substituting Eq.~\eqref{eq:local_stress} into the integral expression in Eq.~\eqref{eq:stress_resultants}, and then separating the mid-surface deformation components $\boldsymbol{\varepsilon}^0$ and $\boldsymbol{\kappa}$ that are independent of the coordinate $z$, leads to the main constitutive equation relating macroscopic resultants to deformation:

\begin{equation}
    \begin{bmatrix} \mathbf{N} \\ \mathbf{M} \end{bmatrix} = 
    \begin{bmatrix} 
        \int \mathbf{Q}(z) dz & \int \mathbf{Q}(z) z dz \\ 
        \int \mathbf{Q}(z) z dz & \int \mathbf{Q}(z) z^2 dz 
    \end{bmatrix} 
    \begin{bmatrix} \boldsymbol{\varepsilon}^0 \\ \boldsymbol{\kappa} \end{bmatrix} =
    \begin{bmatrix} \mathbf{A} & \mathbf{B} \\ \mathbf{B} & \mathbf{D} \end{bmatrix} \begin{bmatrix} \boldsymbol{\varepsilon}^0 \\ \boldsymbol{\kappa} \end{bmatrix}
    \label{eq:abd_derivation}
\end{equation}

The $6 \times 6$ generalized stiffness matrix in Eq.~\eqref{eq:abd_derivation} is called the $\mathbf{ABD}$ matrix, as shown in \textbf{Figure \ref{fig:fig1}b}. It provides a unified basis for evaluating the effective mechanical properties of anisotropic plate/shells. The physical meanings of its submatrices are as follows:

\begin{itemize}
    \item $\mathbf{A}$ \textbf{matrix}, of size $3 \times 3$, is the in-plane membrane stiffness matrix. By definition, $\mathbf{A} = \int \mathbf{Q} dz$ establishes the relation between the in-plane membrane force $\mathbf{N}$ and the mid-surface membrane strain $\boldsymbol{\varepsilon}^0$, and characterizes the overall resistance of the plate/shell to in-plane tension, compression, and shear, as shown in \textbf{Figure \ref{fig:fig1}d}. Here, $A_{11}$ and $A_{22}$ correspond to the in-plane tensile stiffnesses along the two principal directions, $A_{66}$ represents the in-plane shear stiffness, and $A_{12}$ reflects the Poisson-type coupling between orthogonal directions. For isotropic or orthotropic plate/shells, $A_{16}$ and $A_{26}$ are usually zero. In metamaterial structures with directional or chiral topologies, however, these off-diagonal terms may remain nonzero, meaning that axial stretching can induce in-plane shear deformation, i.e., extension-shear coupling.
    
    \item $\mathbf{D}$ \textbf{matrix}, of size $3 \times 3$, is the out-of-plane bending-torsion stiffness matrix. By definition, $\mathbf{D} = \int \mathbf{Q} z^2 dz$ establishes the relation between the bending moment $\mathbf{M}$ and the curvature $\boldsymbol{\kappa}$, and characterizes the resistance of the plate/shell to out-of-plane bending and twisting. Because this matrix involves a second-order integral with respect to the thickness coordinate $z$, its value is sensitive to the distribution of material relative to the neutral surface. Therefore, truncation or loss of surface material in thin-walled Lattice-skin Plate Structures often affects the bending-torsion stiffness, as shown in \textbf{Figure \ref{fig:fig1}f}. Here, $D_{11}$ and $D_{22}$ denote the bending stiffnesses along the two principal directions, $D_{66}$ denotes the out-of-plane torsional stiffness, and $D_{12}$ describes transverse coupling during bending. Nonzero $D_{16}$ and $D_{26}$ correspond to bending-twist coupling, indicating that bending about a single direction may be accompanied by twisting deformation.
    
    \item $\mathbf{B}$ \textbf{matrix}, of size $3 \times 3$, is the extension-bending coupling stiffness matrix. By definition, $\mathbf{B} = \int \mathbf{Q} z dz$ is the first moment of stiffness through the thickness and reflects the coupling between the mid-surface membrane strain $\boldsymbol{\varepsilon}^0$ and the bending moment $\mathbf{M}$, as well as between the curvature $\boldsymbol{\kappa}$ and the membrane force $\mathbf{N}$, as shown in \textbf{Figure \ref{fig:fig1}e}. For plate/shells with a symmetric material distribution through the thickness, the contributions at positive and negative $z$ cancel each other, so the $\mathbf{B}$ matrix vanishes. By contrast, in thin-walled metamaterials with thickness-wise asymmetry, a material distribution offset from the mechanical neutral surface introduces eccentricity effects. As a result, a purely in-plane load may induce macroscopic bending, and pure bending may also be accompanied by a membrane-force response. Here, $B_{11}$, $B_{22}$, and $B_{12}$ mainly describe extension-bending coupling, whereas $B_{16}$ and $B_{26}$ correspond to extension-twist coupling. The latter is particularly important in Lattice-skin Plate Structures with spatial helices, inclined networks, or asymmetric truncation features, and may lead to macroscopic twisting under uniaxial tension.
\end{itemize}

For plate/shells whose stiffness distribution is symmetric about the mid-surface, the contributions at positive and negative thickness coordinates in $\mathbf{B}=\int \mathbf{Q} z dz$ cancel each other, so $\mathbf{B}$ reduces to a zero matrix. For Lattice-skin Plate Structures that contain only a finite number of unit cells through the thickness and are affected by free-surface truncation, however, asymmetry in geometry and material distribution may induce in-plane/out-of-plane coupling responses.

\subsection{LVS-H and static condensation}

LVS-H originates from classical three-dimensional asymptotic homogenization theory and is commonly used to extract the effective mechanical properties of metamaterials, porous media, and composites. This method treats the RVE with complex microstructural geometry as an equivalent solid that is periodically extended in three-dimensional space. In computation, three-dimensional periodic boundary conditions (3D-PBC) are imposed along the three orthogonal directions, including the thickness direction. The microscale fluctuating displacement fields induced by six independent macroscopic strains are then solved, yielding the effective $6 \times 6$ three-dimensional elastic constitutive tensor $\mathbf{C}^H$.

To use these effective 3D solid properties within the macroscopic plate/shell mechanics framework given by Eq.~\eqref{eq:abd_derivation}, an additional reduction in dimensionality is required. A macroscopic thin plate mainly carries in-plane loads and out-of-plane bending. Its upper and lower surfaces are usually free boundaries, so the plane-stress condition can be approximated, namely $\sigma_{33} \approx 0$. Transverse shear stresses, which are only weakly related to the membrane-bending response of the plate/shell, can also be neglected in this framework. Based on these assumptions, the traditional treatment usually employs static condensation to reduce the $6 \times 6$ effective 3D stiffness matrix $\mathbf{C}^H$ to the $3 \times 3$ effective plane-stress stiffness matrix $\mathbf{Q}^H$.

Specifically, under the constraint $\sigma_{33}=0$, the out-of-plane normal strain $\varepsilon_{33}$ can be expressed as a function of the in-plane strain components and substituted back into the 3D constitutive equation, thereby eliminating the normal degree of freedom. The resulting analytical relation between the matrix entries $Q_{ij}^H$ of $\mathbf{Q}^H$ and the effective 3D stiffness components $C_{ij}^H$ can be written as

\begin{equation}
    Q_{ij}^H = C_{ij}^H - \frac{C_{i3}^H C_{j3}^H}{C_{33}^H} \quad (i, j = 1, 2, 6)
    \label{eq:static_condensation}
\end{equation}

After this reduced stiffness matrix has been obtained, the dimensional-reduction procedure in LVS-H substitutes it into the macroscopic integral formulas derived in the previous subsection. At this stage, the microscopic metamaterial is assumed, at the macroscopic level, to be a homogeneous continuum uniformly smeared through the thickness. Analytical integration of the matrix $\mathbf{Q}^H$ over the physical plate thickness $h$ then gives the effective in-plane, extension-bending coupling, and bending stiffness matrices:

\begin{equation}
    (\mathbf{A}, \mathbf{B}, \mathbf{D}) = \int_{-h/2}^{h/2} \mathbf{Q}^H (1, z, z^2) dz = \left( \mathbf{Q}^H h, \; \mathbf{0}, \; \mathbf{Q}^H \frac{h^3}{12} \right)
    \label{eq:traditional_integration}
\end{equation}

This treatment, which first extracts effective 3D material properties and then introduces them into the plate/shell constitutive integration, is computationally efficient and convenient to implement, and is therefore common in routine engineering analysis. However, it implicitly assumes periodic continuation through the thickness and uses the effective properties as parameters of a continuous homogeneous material in the macroscopic plate/shell integration. For Lattice-skin Plate Structures that contain only a finite number of unit cells through the thickness, this assumption may deviate from the actual boundary state of the structure. Specifically, the 3D-PBC imposed through the thickness constrains the microscopic deformation modes of the upper and lower free surfaces, making it difficult to capture the free-surface effect caused by physical truncation.

\subsection{LPS-H framework for finite-thickness evaluation}
To reduce the mechanical bias introduced by the thickness-wise periodic continuation assumption in LVS-H, especially the stiffness variation and sectional asymmetry caused by truncation of a finite number of unit cells, LPS-H is introduced for finite-thickness plate/shell structures. Starting from the microscopic boundary-value problem, this method considers both the geometric features of the plate/shell and the actual boundary state during homogenization, thereby enabling dimensional reduction from a 3D heterogeneous solid to a 2D equivalent plate/shell model.

In the finite element formulation of LPS-H, the boundary conditions of the RVE are closer to the service boundary state of a thin-walled plate/shell. The method imposes two-dimensional periodic boundary conditions (2D-PBC) only in the in-plane $x$ and $y$ directions to represent periodic continuation in plane. In the out-of-plane thickness direction $z$, the model retains one-dimensional free boundary conditions (1D-FBC), thereby releasing the kinematic constraints on the upper and lower surfaces. These mixed boundary conditions allow thickness-wise Poisson deformation and local warping under loading, which reduces the stiffness overestimation caused by fully three-dimensional periodic constraints.

Under these mixed boundary conditions, LPS-H extracts the generalized plate/shell stiffness matrix through a finite element algebraic formulation. The procedure can be summarized in the following four steps:

\subsubsection*{Step 1: The mathematical strategy of column-by-column probing}

According to Eq.~\eqref{eq:abd_derivation}, the unknown $\mathbf{ABD}$ stiffness matrix is a $6 \times 6$ linear mapping operator. It maps the 6-dimensional generalized mid-surface deformation vector $\mathbf{E} = [\boldsymbol{\varepsilon}^0, \boldsymbol{\kappa}]^T$ to the 6-dimensional generalized internal-force response vector $\mathbf{R} = [\mathbf{N}, \mathbf{M}]^T$.

In numerical homogenization, the $\mathbf{ABD}$ matrix is unknown, so a column-by-column probing strategy is adopted. According to the rule of matrix multiplication, if an orthogonal unit deformation basis vector $\mathbf{E}^{(I)}$ is applied such that only its $I$th component is equal to 1 and all other components are zero, then the resulting internal-force response vector $\mathbf{R}^{(I)}$ is numerically equal to the $I$th column of the $\mathbf{ABD}$ matrix. Therefore, solving the full $6 \times 6$ matrix can be converted into solving six independent microscopic response problems under six macroscopic unit deformation modes, where $I = 1, 2, \dots, 6$.

\subsubsection*{Step 2: Construction of the macroscopic unit deformation modes}

To transfer the above 6-dimensional macroscopic unit deformation vector $\mathbf{E}^{(I)}$ to the 3D RVE, it must be decomposed and mapped to the initial local strain at any point inside the RVE. The vector $\mathbf{E}^{(I)}$ can be split into an in-plane membrane part $\boldsymbol{\Lambda}_{mem}^{(I)}$ and an out-of-plane bending part $\boldsymbol{\Lambda}_{bend}^{(I)}$, corresponding to the first three and last three components, respectively:

\begin{itemize}
    \item \textbf{Cases $I = 1, 2, 3$: in-plane membrane strain modes.} Only in-plane tension and shear are activated.\\
    The membrane components take the basis vectors $\boldsymbol{\Lambda}_{mem}^{(1)} = [1, 0, 0]^T$, $\boldsymbol{\Lambda}_{mem}^{(2)} = [0, 1, 0]^T$, and $\boldsymbol{\Lambda}_{mem}^{(3)} = [0, 0, 1]^T$.\\
    The bending components remain zero: $\boldsymbol{\Lambda}_{bend}^{(1,2,3)} = [0, 0, 0]^T$.
    
    \item \textbf{Cases $I = 4, 5, 6$: out-of-plane bending and twisting curvature modes.} Only out-of-plane bending and twisting are activated.\\
    The membrane components remain zero: $\boldsymbol{\Lambda}_{mem}^{(4,5,6)} = [0, 0, 0]^T$.\\
    The bending components take the basis vectors $\boldsymbol{\Lambda}_{bend}^{(4)} = [1, 0, 0]^T$, $\boldsymbol{\Lambda}_{bend}^{(5)} = [0, 1, 0]^T$, and $\boldsymbol{\Lambda}_{bend}^{(6)} = [0, 0, 1]^T$.
\end{itemize}

Under the plane-section assumption of plate/shell theory, the initial macroscopic strain field $\boldsymbol{\varepsilon}^{macro(I)}(z)$ induced at the point whose normal coordinate from the geometric mid-surface is $z$ inside the RVE can be written in the unified form

\begin{equation}
    \boldsymbol{\varepsilon}^{macro(I)}(z) = \boldsymbol{\Lambda}_{mem}^{(I)} + z \boldsymbol{\Lambda}_{bend}^{(I)}
    \label{eq:macro_strain_modes}
\end{equation}

\subsubsection*{Step 3: Stress recovery under mixed boundary conditions}

After Eq.~\eqref{eq:macro_strain_modes} is applied to the RVE, the initial macroscopic strain field disturbs the local equilibrium of the microscopic solid because the metamaterial contains an interwoven topology of solid matrix and voids. To re-establish equilibrium, the structure develops a local fluctuating displacement field, together with the fluctuating strain $\boldsymbol{\varepsilon}^{fluc(I)}(\mathbf{x})$. By solving the finite element weak form under the mixed boundary conditions of 2D-PBC and 1D-FBC, this fluctuating strain can be obtained, and the microscopic local in-plane stress field $\boldsymbol{\sigma}_p^{(I)}(\mathbf{x})$ for the $I$th loading case can be recovered from the local constitutive tensor $\mathbf{C}(\mathbf{x})$ of the solid material:

\begin{equation}
    \boldsymbol{\sigma}_p^{(I)}(\mathbf{x}) = \mathbf{C}(\mathbf{x}) \left[ \boldsymbol{\varepsilon}^{macro(I)}(z) + \boldsymbol{\varepsilon}^{fluc(I)}(\mathbf{x}) \right]
    \label{eq:true_micro_stress}
\end{equation}

\subsubsection*{Step 4: Area integration and ABD matrix extraction}

After obtaining the microscopic stress fields for the six loading cases, LPS-H returns to the physical definition of the macroscopic internal resultants. By taking spatial moment integrals of $\boldsymbol{\sigma}_p^{(I)}(\mathbf{x})$ over the microscopic solid volume domain $\Omega$ and dividing by the macroscopic projected area $S$ of the RVE, the macroscopic generalized internal-force vector $\mathbf{R}^{(I)}$ is obtained:

\begin{equation}
    \begin{bmatrix} \mathbf{N}^{(I)} \\ \mathbf{M}^{(I)} \end{bmatrix} = \frac{1}{S} \int_{\Omega} \begin{bmatrix} \boldsymbol{\sigma}_p^{(I)}(\mathbf{x}) \\ \boldsymbol{\sigma}_p^{(I)}(\mathbf{x}) z \end{bmatrix} d\Omega
    \label{eq:extract_column}
\end{equation}

As described in Step 1, the resulting internal-force vector corresponds numerically to the $I$th column of the $\mathbf{ABD}$ matrix. In this way, a mapping is established from microscopic stress integration to macroscopic stiffness components:

\begin{itemize}
    \item \textbf{When $I = 1, 2, 3$, corresponding to in-plane tension and shear cases:}\\
    the in-plane membrane force $\mathbf{N}^{(I)}$ obtained from the upper part of the integral expression forms Columns 1, 2, and 3 of the $\mathbf{A}$ matrix;\\
    the out-of-plane bending moment $\mathbf{M}^{(I)}$ obtained from the lower part forms Columns 1, 2, and 3 of the extension-bending coupling matrix $\mathbf{B}$.
    
    \item \textbf{When $I = 4, 5, 6$, corresponding to out-of-plane bending and twisting cases:}\\
    the in-plane membrane force $\mathbf{N}^{(I)}$ obtained from the upper part of the integral expression forms Columns 4, 5, and 6 of the coupling matrix $\mathbf{B}$; according to the reciprocity theorem, this block is theoretically symmetric with the $\mathbf{B}$ block obtained in the previous step;\\
    the out-of-plane bending moment $\mathbf{M}^{(I)}$ obtained from the lower part forms Columns 1, 2, and 3 of the bending stiffness matrix $\mathbf{D}$.
\end{itemize}

\section{Program Structure and Python Implementation}
The compact Python open-source code developed for LPS-H is analyzed below. It relies on the standard scientific computing libraries \texttt{NumPy} and \texttt{SciPy}, together with the \texttt{CuPy} library for GPU heterogeneous acceleration. The full program is modularized into four main functions. Following the execution flow of plate/shell finite element homogenization, the main code blocks and their underlying vectorized algorithms are described. The code is available in the open-source repository at \href{https://github.com/TopJournals/ThinWalledHomogenization/blob/main/core/plate_homogenizer.py}{\texttt{core/plate\_homogenizer.py}}.

\subsection{Material Definition and Element Stiffness}

The first step of the homogenization calculation is to define the constitutive relation of the microscopic base material and the element stiffness matrix of the discretized mesh. In this code, these tasks are implemented by the two functions \texttt{get\_isotropic\_elasticity} and \texttt{compute\_element\_stiffness}.

\begin{lstlisting}[language=Python, numbers=left, firstnumber=8]
def get_isotropic_elasticity(E=1.0, nu=0.3):
    lam, mu = E * nu / ((1 + nu) * (1 - 2 * nu)), E / (2 * (1 + nu))  # Calculate Lame parameters
    C = np.zeros((6, 6))  # Initialize 6x6 constitutive matrix
    C[0:3, 0:3] = lam
    np.fill_diagonal(C[0:3, 0:3], lam + 2 * mu)
    np.fill_diagonal(C[3:6, 3:6], mu)
    return C
\end{lstlisting}

As shown in Lines 8-14, the program first computes the Lam\'e constants $\lambda$ and $\mu$ from the input Young's modulus $E$ and Poisson's ratio $\nu$.
\begin{equation}
    \label{eq:Lamé}
    \lambda = \frac{\nu E} {(1+\nu)(1-2\nu)}, \mu = \frac E {2(1+\nu)}
\end{equation}

The code then uses \texttt{NumPy} array slicing and the diagonal-filling function \texttt{fill\_diagonal} to construct the $6 \times 6$ constitutive matrix $\mathbf{C}$ for an isotropic linear elastic material.
\begin{equation}
    \label{eq:elastic tensor}
    \boldsymbol{C} = \lambda \begin{bmatrix}
        1 & 1 & 1 & 0 & 0 & 0 \\
        1 & 1 & 1 & 0 & 0 & 0 \\
        1 & 1 & 1 & 0 & 0 & 0 \\
        0 & 0 & 0 & 0 & 0 & 0 \\
        0 & 0 & 0 & 0 & 0 & 0 \\
        0 & 0 & 0 & 0 & 0 & 0
    \end{bmatrix}
    + \mu \begin{bmatrix}
        2 & 0 & 0 & 0 & 0 & 0 \\
        0 & 2 & 0 & 0 & 0 & 0 \\
        0 & 0 & 2 & 0 & 0 & 0 \\
        0 & 0 & 0 & 1 & 0 & 0 \\
        0 & 0 & 0 & 0 & 1 & 0 \\
        0 & 0 & 0 & 0 & 0 & 1
    \end{bmatrix}
\end{equation}

After the material constitutive tensor has been obtained, the code discretizes the metamaterial RVE using standard eight-node hexahedral elements.

\begin{lstlisting}[language=Python, numbers=left, firstnumber=17]
def compute_element_stiffness(C, dx, dy, dz):
    Ke, Bs = np.zeros((24, 24)), []
    nodes = np.array([[-1, -1, -1], [1, -1, -1], [1, 1, -1], [-1, 1, -1], [-1, -1, 1], [1, -1, 1], [1, 1, 1], [-1, 1, 1]])  # 8-node local coordinates
    detJ = dx * dy * dz / 8.0
    for q in nodes / np.sqrt(3):  # Loop over 8 Gauss integration points
        dN = 0.125 * np.array([(1 + q[1] * nodes[:, 1]) * (1 + q[2] * nodes[:, 2]) * nodes[:, 0] * (2 / dx), (1 + q[0] * nodes[:, 0]) * (1 + q[2] * nodes[:, 2]) * nodes[:, 1] * (2 / dy), (1 + q[0] * nodes[:, 0]) * (1 + q[1] * nodes[:, 1]) * nodes[:, 2] * (2 / dz)])  # Cartesian derivatives
        B = np.zeros((6, 24))  # Strain-displacement matrix
        B[0, 0::3], B[1, 1::3], B[2, 2::3] = dN[0], dN[1], dN[2]
        B[3, 1::3], B[3, 2::3] = dN[2], dN[1]
        B[4, 0::3], B[4, 2::3] = dN[2], dN[0]
        B[5, 0::3], B[5, 1::3] = dN[1], dN[0]
        Bs.append(B)  # Store B matrices for stress recovery
        Ke += B.T @ C @ B * detJ  # Accumulate local stiffness
    return Ke, Bs, detJ
\end{lstlisting}

Lines 17-30 show the analytical calculation of the element stiffness matrix $\mathbf{K}_e$. To evaluate the stiffness matrix on a regular voxel grid, the program introduces an isoparametric mapping from the physical coordinate system $(x, y, z)$ to the dimensionless local natural coordinates $(\xi, \eta, \zeta \in [-1, 1])$. For an orthogonal regular element with side lengths $dx$, $dy$, and $dz$, the Jacobian matrix $\mathbf{J}$ degenerates into a diagonal matrix, and its determinant is constant over the entire element domain:
\begin{equation}
    |J| = \det(\mathbf{J}) = \frac{dx}{2} \cdot \frac{dy}{2} \cdot \frac{dz}{2} = \frac{dx \cdot dy \cdot dz}{8}
\end{equation}

The theoretical definition of the stiffness matrix is the volume integral over the physical domain, $\mathbf{K}_e = \int_{V_e} \mathbf{B}^T \mathbf{C} \mathbf{B} dV$. The code uses a $2 \times 2 \times 2$ Gauss-Legendre quadrature rule to convert this volume integral into a discrete sum over eight integration points:
\begin{equation}
    \mathbf{K}_e = \sum_{g=1}^{8} \mathbf{B}^T(\mathbf{q}_g) \mathbf{C} \mathbf{B}(\mathbf{q}_g) |J| W_g
\end{equation}
Because the Gauss-point coordinates in each direction are $\pm 1/\sqrt{3}$ and the corresponding integration weights satisfy $w_i = 1$, the total 3D weight is $W_g = 1 \times 1 \times 1 = 1$. Here, $\mathbf{q}_g = (\xi_g, \eta_g, \zeta_g)$ denotes the coordinates of the eight Gauss points.

For an eight-node hexahedral element, the shape function $N_i$ of Node $i$ is defined as
\begin{equation}
    N_i(\xi, \eta, \zeta) = \frac{1}{8} (1 + \xi_i \xi) (1 + \eta_i \eta) (1 + \zeta_i \zeta) \quad (i = 1, \dots, 8)
\end{equation}

To construct the strain-displacement matrix $\mathbf{B}$, the derivatives of the shape functions with respect to the physical coordinates $(x,y,z)$ must be computed, which can be written as
\begin{equation}
    \frac{\partial N_i}{\partial x} = \frac{\partial N_i}{\partial \xi} \frac{\partial \xi}{\partial x}
\end{equation}

Because the mapping is linear, the geometric scaling factor is $\frac{\partial \xi}{\partial x} = \frac{2}{dx}$. The array \texttt{dN} in Line 22 uses vectorized operations to compute the physical-space derivatives of all eight nodes at the current integration point simultaneously:
\begin{equation}
    \begin{bmatrix} \texttt{dN[0]} \\ \texttt{dN[1]} \\ \texttt{dN[2]} \end{bmatrix} = 
    \begin{bmatrix} \frac{\partial \mathbf{N}}{\partial x} \\ \frac{\partial \mathbf{N}}{\partial y} \\ \frac{\partial \mathbf{N}}{\partial z} \end{bmatrix} = 
    \frac{1}{8}
    \begin{bmatrix} 
        \xi_i (1 + \eta_i \eta_g) (1 + \zeta_i \zeta_g) \frac{2}{dx} \\
        \eta_i (1 + \xi_i \xi_g) (1 + \zeta_i \zeta_g) \frac{2}{dy} \\
        \zeta_i (1 + \xi_i \xi_g) (1 + \eta_i \eta_g) \frac{2}{dz}
    \end{bmatrix}_{1 \times 8}
\end{equation}

Next, Lines 23-28 arrange $\varepsilon_{xx}$, $\varepsilon_{yy}$, $\varepsilon_{zz}$, $\gamma_{yz}$, $\gamma_{xz}$, and $\gamma_{xy}$ according to Voigt notation and assemble the derivative vectors into the $6 \times 24$ strain-displacement matrix $\mathbf{B}$:
\begin{equation}
    \mathbf{B} = 
    \begin{bmatrix}
        \frac{\partial N_1}{\partial x} & 0 & 0 & \dots & \frac{\partial N_8}{\partial x} & 0 & 0 \\
        0 & \frac{\partial N_1}{\partial y} & 0 & \dots & 0 & \frac{\partial N_8}{\partial y} & 0 \\
        0 & 0 & \frac{\partial N_1}{\partial z} & \dots & 0 & 0 & \frac{\partial N_8}{\partial z} \\
        0 & \frac{\partial N_1}{\partial z} & \frac{\partial N_1}{\partial y} & \dots & 0 & \frac{\partial N_8}{\partial z} & \frac{\partial N_8}{\partial y} \\
        \frac{\partial N_1}{\partial z} & 0 & \frac{\partial N_1}{\partial x} & \dots & \frac{\partial N_8}{\partial z} & 0 & \frac{\partial N_8}{\partial x} \\
        \frac{\partial N_1}{\partial y} & \frac{\partial N_1}{\partial x} & 0 & \dots & \frac{\partial N_8}{\partial y} & \frac{\partial N_8}{\partial x} & 0
    \end{bmatrix}
\end{equation}
In Line 29, the chained matrix multiplication \texttt{B.T @ C @ B * detJ} accumulates the stiffness contribution from a single integration point. At the same time, the $\mathbf{B}$ matrix at each Gauss point is stored in the list \texttt{Bs}, which preserves the required geometric mapping operators for recovering local fluctuation stresses and computing the $\mathbf{D}$ matrix and the lower coupling block in the post-processing stage.

\subsection{Tensor-based Degree-of-Freedom (DOF) Mapping and Mixed Boundaries}

\begin{figure*}[htbp]
  \centering
  \includegraphics[width=\textwidth]{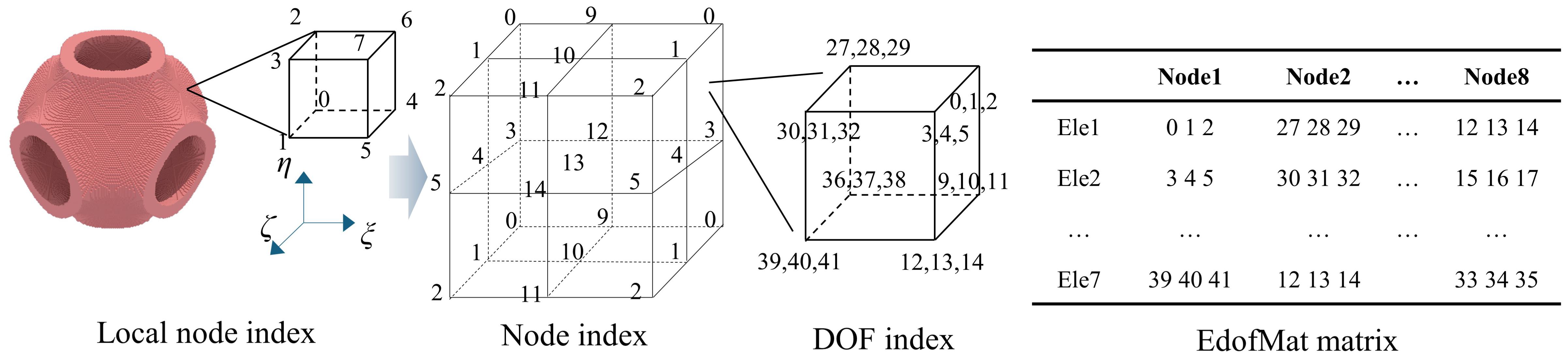}
  \caption{Schematic of the loop-free assembly logic for the local-to-global DOF mapping tensor \texttt{edofMat} based on multidimensional tensor slicing.}
  \label{fig:tensor_assembly}
\end{figure*}

In conventional high-fidelity finite element programs, determining the mapping between local element degrees of freedom (DOFs) and global system DOFs is often one of the most time-consuming steps. This is especially true when periodic boundary conditions (PBCs) must be handled in metamaterial models. Traditional codes usually rely on multiple nested \texttt{for} loops and node-coordinate matching with geometric tolerances, so the computational cost increases with mesh resolution. To address this issue, the function \texttt{build\_tensor\_dof\_mapping} introduces a loop-free mapping algorithm that accelerates the process using \texttt{NumPy}-based tensor vectorization, as illustrated in \textbf{Figure \ref{fig:tensor_assembly}}.

\begin{lstlisting}[language=Python, numbers=left, firstnumber=33]
def build_tensor_dof_mapping(voxel, dx, dy, dz, thickness):
    nx, ny, nz = voxel.shape  # Enforce XYZ indexing convention
    node_indices = np.arange((nx + 1) * (ny + 1) * (nz + 1)).reshape(nx + 1, ny + 1, nz + 1)
    dof_tensor = np.zeros((nx + 1, ny + 1, nz + 1, 3), dtype=int)
    dof_tensor[..., 0], dof_tensor[..., 1], dof_tensor[..., 2] = node_indices * 3, node_indices * 3 + 1, node_indices * 3 + 2
    dof_tensor[nx, :, :, :] = dof_tensor[0, :, :, :]
    dof_tensor[:, ny, :, :] = dof_tensor[:, 0, :, :]

    n1, n2 = dof_tensor[:-1, :-1, :-1, :], dof_tensor[1:, :-1, :-1, :]
    n3, n4 = dof_tensor[1:, 1:, :-1, :], dof_tensor[:-1, 1:, :-1, :]
    n5, n6 = dof_tensor[:-1, :-1, 1:, :], dof_tensor[1:, :-1, 1:, :]
    n7, n8 = dof_tensor[1:, 1:, 1:, :], dof_tensor[:-1, 1:, 1:, :]
    edof_tensor = np.concatenate([n1, n2, n3, n4, n5, n6, n7, n8], axis=-1)

    active_mask = voxel > 0  # Filter out void regions
    edofMat = edof_tensor[active_mask]

    z_coords = np.linspace(dz / 2, thickness - dz / 2, nz) - thickness / 2.0
    z_grid = np.broadcast_to(z_coords[None, None, :], (nx, ny, nz))  # Z is the 3rd axis
    z_active = z_grid[active_mask]
    return edofMat, z_active, node_indices.size * 3
\end{lstlisting}

This code block constructs the local-to-global DOF mapping and imposes the mixed boundary conditions required by LPS-H without introducing explicit loops.

In Lines 34-37, the nodes of the structured voxel grid are arranged regularly in space. To match the Cartesian coordinate system used in mechanics, the program adopts the indexing order $x$-$y$-$z$. Line 35 uses \texttt{np.arange} to generate a sequential node-index tensor for the discrete space. Line 36 then builds the four-dimensional global DOF tensor \texttt{dof\_tensor} with shape \texttt{(nx+1, ny+1, nz+1, 3)}. In this tensor, the first three dimensions correspond to the 3D physical positions $(x, y, z)$ of the nodes, while the last dimension stores the three orthogonal physical DOFs $u$, $v$, and $w$ associated with each node.

In a traditional finite element formulation, the 2D-PBC requires the nodal displacements on opposite boundaries to be identical, namely $\mathbf{u}(L_x, y, z) = \mathbf{u}(0, y, z)$ and $\mathbf{u}(x, L_y, z) = \mathbf{u}(x, 0, z)$. These kinematic constraints are usually introduced by Lagrange multipliers or penalty methods, which increase both system size and solution complexity. By contrast, Lines 38-39 use in-place memory slicing of the global DOF tensor to overwrite the DOF numbers of the slave nodes on the $X=L_x$ and $Y=L_y$ boundaries with the corresponding master-node numbers on the negative sides. The $z$ axis, which represents the thickness direction, is left unchanged. In this way, periodic mapping is enforced in the in-plane $x$ and $y$ directions while the free boundary conditions on the upper and lower surfaces are retained.

To obtain the nodal assembly relation inside each hexahedral element, Lines 41-44 use shifted array slices in different directions. For example, \texttt{[1:, :-1, :-1]} extracts the tensor \texttt{n2}, which corresponds to Node 2 after a positive shift along the $x$ direction. Through this combination of data-structure slicing and \texttt{np.concatenate}, the DOFs of the eight corner nodes of each voxel element are extracted in a single step, and the $N_{elem} \times 24$ DOF mapping tensor \texttt{edof\_tensor} for the entire computational domain is constructed. The coordinate ordering is kept consistent with the related educational paper\citep{andreassenHowDetermineComposite2014} to improve code portability.

Because metamaterials contain many topological voids, removing void elements reduces the equilibrium system and accelerates the solution. Lines 47-48 use the Boolean mask \texttt{voxel > 0} generated from the voxel model to filter out the material-free regions and obtain the compact mapping matrix \texttt{edofMat} for the active solid elements.

Furthermore, in LPS-H, evaluation of the bending stiffness matrix $\mathbf{D}$ and the extension-bending coupling matrix $\mathbf{B}$ depends on the material distribution through the thickness. Lines 50-52 explicitly extract the one-dimensional vector \texttt{z\_active}. Through the coordinate transformation $z = \bar{z} - h/2$, this vector computes and stores the normal thickness coordinate of the centroid of each active solid element relative to the geometric mid-surface of the macroscopic plate/shell.

\subsection{Loop-Free Assembly of Global Matrices}

In high-resolution voxel-based finite element analysis, assembling the global stiffness matrix $\mathbf{K}$ and load matrix $\mathbf{F}$ with traditional nested loops leads to substantial time overhead. To address this issue, this code block uses the low-level vectorization capability of \texttt{NumPy} to realize loop-free matrix assembly.

\begin{lstlisting}[language=Python, numbers=left, firstnumber=68]
    iK, jK = np.repeat(edofMat, 24, axis=1).flatten(), np.tile(edofMat, (1, 24)).flatten()
    sK = np.tile(Ke.flatten(), edofMat.shape[0])
    K = coo_matrix((sK, (iK, jK)), shape=(total_dofs, total_dofs)).tocsr()  # Loop-free assembly of global stiffness

    E_macro = np.zeros((len(z_active), 6, 6))
    E_macro[:, 0, 0], E_macro[:, 1, 1], E_macro[:, 5, 2] = 1.0, 1.0, 1.0  # Apply unit membrane strains
    E_macro[:, 0, 3], E_macro[:, 1, 4], E_macro[:, 5, 5] = z_active, z_active, z_active  # Apply unit bending curvatures

    F_ele = sum([np.einsum('ji,kjl->kil', Bs[i], np.einsum('ij,kjl->kil', C, E_macro)) * detJ for i in range(8)])  # Local load via tensor contraction
    F = np.column_stack([np.bincount(edofMat.flatten(), weights=F_ele[:, :, c].flatten(), minlength=total_dofs) for c in range(6)])  # Global load assembly
\end{lstlisting}

The global stiffness matrix $\mathbf{K}$ is obtained by assembling the local element stiffness matrices $\mathbf{K}_e$ according to the topological mapping matrix \texttt{edofMat}. The traditional approach uses \texttt{for} loops to add each $24 \times 24$ submatrix to the global matrix one by one. Lines 68-70 instead exploit the properties of the coordinate (COO) sparse matrix format to replace explicit loops with data operations. Through \texttt{np.repeat} and \texttt{np.tile}, the program tiles the local mapping tensor \texttt{edofMat} along different dimensions, thereby generating the global one-dimensional row-index vector \texttt{iK} and column-index vector \texttt{jK}, each of length $N_{active} \times 24^2$. At the same time, the element stiffness values are flattened into the one-dimensional vector \texttt{sK}. Line 70 then calls \texttt{scipy.sparse.coo\_matrix}, which performs memory allocation and accumulation for millions of nonzero entries at the underlying C level, and finally converts the result into compressed sparse row (CSR) format for better compatibility.

Before the global load vector is assembled, the six macroscopic unit deformation modes in LPS-H must be converted into local strain distributions at the microscopic scale. Line 72 initializes the three-dimensional loading super-tensor \texttt{E\_macro} with shape $(N_{active}, 6, 6)$. Its first dimension corresponds to the active solid elements, its second dimension corresponds to the six components of the local strain, and its third dimension corresponds to the six independent loading cases, namely three in-plane membrane strains and three out-of-plane curvatures. In the 3D finite element formulation, the local strain vector of each microscopic solid element follows Voigt notation. Accordingly, the row indices $0 \sim 5$ of \texttt{E\_macro} represent

\begin{equation}
    \boldsymbol{\varepsilon}^{3D} = [\varepsilon_{11}, \varepsilon_{22}, \varepsilon_{33}, \gamma_{23}, \gamma_{13}, \gamma_{12}]^T
\end{equation}

In LPS-H, by contrast, six independent macroscopic deformation cases are probed in sequence. These correspond to the column indices $0 \sim 5$ of \texttt{E\_macro}, and the associated deformation basis vector is

\begin{equation}
    \mathbf{E} = [\varepsilon_{11}^0, \varepsilon_{22}^0, \gamma_{12}^0, \kappa_{11}, \kappa_{22}, \kappa_{12}]^T
\end{equation}

According to the plane-section assumption for plates, $\boldsymbol{\varepsilon} = \boldsymbol{\varepsilon}^0 + z\boldsymbol{\kappa}$. For any 3D solid element located at the normal coordinate $z$ from the mid-surface, the initial local strain induced under the six macroscopic loading cases can be expressed explicitly through a $6 \times 6$ transformation matrix $\boldsymbol{\Gamma}(z)$:

\begin{equation}
    \label{eq:macro_mapping_matrix}
    \begin{aligned}
        % First line: left-hand side of the original expression
        \boldsymbol{\varepsilon}^{3D}(z) &= \boldsymbol{\Gamma}(z) \mathbf{I}_{6 \times 6} \\
        % Second line: 2x2 grid layout
        &= \begin{array}[b]{c@{\quad}l}
            % Left side of the first grid row: column labels
            \begin{matrix}
                % \text{\scriptsize Col.\ 0} & \text{\scriptsize Col.\ 1} & \text{\scriptsize Col.\ 2} & \text{\scriptsize Col.\ 3} & \text{\scriptsize Col.\ 4} & \text{\scriptsize Col.\ 5} \\
                \varepsilon_{11}^0 & \varepsilon_{22}^0 & \gamma_{12}^0 & \kappa_{11} & \kappa_{22} & \kappa_{12} \\
                \downarrow & \downarrow & \downarrow & \downarrow & \downarrow & \downarrow
            \end{matrix} & \\[-1mm] % Leave the right side empty and adjust the spacing to the matrix below
            
            % Left side of the second grid row: matrix body
            \begin{bmatrix}
                \ 1\quad & 0\quad & 0\quad & z\quad & 0\quad & 0 \ \\
                \ 0\quad & 1\quad & 0\quad & 0\quad & z\quad & 0 \ \\
                \ 0\quad & 0\quad & 0\quad & 0\quad & 0\quad & 0 \ \\
                \ 0\quad & 0\quad & 0\quad & 0\quad & 0\quad & 0 \ \\
                \ 0\quad & 0\quad & 0\quad & 0\quad & 0\quad & 0 \ \\
                \ 0\quad & 0\quad & 1\quad & 0\quad & 0\quad & z \ 
            \end{bmatrix} &
            % Right side of the second grid row: row labels
            \begin{matrix}
                \leftarrow \varepsilon_{11} \\
                \leftarrow \varepsilon_{22} \\
                \leftarrow \varepsilon_{33} \\
                \leftarrow \gamma_{23} \\
                \leftarrow \gamma_{13} \\
                \leftarrow \gamma_{12}
            \end{matrix}
        \end{array}
    \end{aligned}
\end{equation}

The construction of \texttt{E\_macro} bridges 3D continuum mechanics and 2D plate/shell mechanics, while maintaining consistency in the application of external loading without cumbersome loops.

For Element $e$, under the $I$th deformation case, the equivalent nodal load induced by the initial macroscopic strain is written as
\begin{equation}
    \mathbf{F}_{e}^{(I)} = \int_{V_e} \mathbf{B}^T \mathbf{C} \boldsymbol{\varepsilon}^{macro(I)} dV
\end{equation}

If the traditional approach were used, evaluating this set of loads for all elements under all six loading cases would require multiple nested loops. Line 76 introduces the Einstein summation function \texttt{np.einsum}. The inner contraction, \texttt{einsum('ij,kjl->kil', C, E\_macro)}, computes the initial stresses of the solid elements under the six loading cases. The outer contraction then combines these stresses with the strain-displacement matrix \texttt{Bs[i]} at each of the eight Gauss points through loop-free tensor contraction. Because the Jacobian determinant is constant, the program multiplies directly by the integration weight \texttt{detJ} outside the contraction to obtain the local $24 \times 6$ load array \texttt{F\_ele}. Then, in Line 77, the one-dimensional counting function \texttt{np.bincount} is used to assemble the local load array into the six right-hand-side column vectors of the global system. Here, \texttt{bincount} treats the global DOF numbers in \texttt{edofMat} as index bins and accumulates the local nodal forces into the global load matrix $\mathbf{F}$ using \texttt{F\_ele} as weights.

\subsection{Asynchronous GPU Solving via CuPy}

The computational bottleneck of the homogenization procedure is often the solution of the large sparse linear system $\mathbf{K}\mathbf{U} = \mathbf{F}$. To improve hardware utilization, the code uses the \texttt{CuPy} library to implement heterogeneous acceleration between the central processing unit (CPU) and the graphics processing unit (GPU), together with a non-blocking concurrent-stream mechanism.

\begin{lstlisting}[language=Python, numbers=left, firstnumber=78]
    active_dofs = np.setdiff1d(np.unique(edofMat), np.unique(edofMat)[:3])  # Eliminate rigid body motions

    U = np.zeros((total_dofs, 6))
    K_active_gpu = cpsp.csr_matrix(K[active_dofs, :][:, active_dofs])
    F_active_gpu = cp.asarray(F[active_dofs, :])
    U_active_gpu = cp.zeros((len(active_dofs), 6))
    M_gpu = cpsp.diags(1.0 / K_active_gpu.diagonal())
    streams = [cp.cuda.Stream(non_blocking=True) for _ in range(6)]  # Initialize concurrent CUDA streams
    for c in range(6):
        with streams[c]:  # Launch asynchronous GPU solving for 6 load cases
            U_col_gpu, _ = cpspla.cg(K_active_gpu, F_active_gpu[:, c], M=M_gpu, tol=1e-6, maxiter=5000)  # Jacobi Preconditioner Conjugate gradient solver on GPU
            U_active_gpu[:, c] = U_col_gpu
    for stream in streams:
        stream.synchronize()  # Barrier synchronization
    U[active_dofs, :] = U_active_gpu.get()
\end{lstlisting}

Before solving the system, rigid-body motion must be removed. As shown in Line 78, the code uses \texttt{np.setdiff1d} to eliminate the first three DOFs as reference anchors and extract the nonsingular active DOF set \texttt{active\_dofs}. In Lines 81-82, the stiffness and load matrices are converted to \texttt{CuPy} format and transferred to NVIDIA GPU memory. To increase the throughput of the GPU streaming multiprocessors, Line 83 instantiates six non-blocking Compute Unified Device Architecture (CUDA) streams. In the following loop, six independent Jacobi-preconditioned conjugate gradient (PCG) solves using \texttt{cpspla.cg} are dispatched to the GPU concurrently. Finally, the barrier operation \texttt{stream.synchronize()} ensures that all calculations are complete, and the results $\mathbf{U}$ are transferred back to host memory through the \texttt{.get()} method.

The performance of this solver depends on CUDA-capable NVIDIA GPU hardware together with the \texttt{CuPy} heterogeneous computing library. To preserve the generality and portability of the open-source code, the original GPU solver module can be replaced by the standard sparse solver provided by \texttt{SciPy} when the local environment supports only CPU computation.

The CPU-only fallback code is shown below:

\begin{lstlisting}[language=Python, numbers=left, firstnumber=78]
    active_dofs = np.setdiff1d(np.unique(edofMat), np.unique(edofMat)[:3])  # Eliminate rigid body motions

    U = np.zeros((total_dofs, 6))
    # Extract non-singular active matrices on CPU
    K_active = K[active_dofs, :][:, active_dofs]
    F_active = F[active_dofs, :]
    U_active = np.zeros((len(active_dofs), 6))
    import scipy.sparse.linalg as spla
    for c in range(6):  # Sequential solving for 6 load cases on CPU
        U_col, _ = spla.cg(K_active, F_active[:, c])  # SciPy Conjugate gradient solver
        U_active[:, c] = U_col
    U[active_dofs, :] = U_active
\end{lstlisting}

\subsection{Area Integration for ABD Matrix}

After solving for the microscopic fluctuation displacement $\mathbf{U}$, the program recovers the physical stress field and extracts the $\mathbf{ABD}$ stiffness matrix through macroscopic area integration.

\begin{lstlisting}[language=Python, numbers=left, firstnumber=94]
    ABD = np.zeros((6, 6))
    for i_gp in range(8):
        Sigma = np.einsum('ij,kjl->kil', C, E_macro - np.einsum('ij,kjl->kil', Bs[i_gp], U[edofMat, :]))  # Recover true physical stress
        ABD[0:3, :] += np.sum(Sigma[:, [0, 1, 5], :], axis=0) * detJ / plate_area  # Extract A and B matrices (0th moment)
        ABD[3:6, :] += np.sum(Sigma[:, [0, 1, 5], :] * z_active[:, None, None], axis=0) * detJ / plate_area  # Extract D and B* matrices (1st moment)
    return (ABD + ABD.T) / 2.0  # Symmetrization for Maxwell-Betti reciprocity
\end{lstlisting}

In a 3D heterogeneous microstructure, the local strain is the superposition of the imposed macroscopic initial strain field and the microscopic fluctuation strain caused by the pore topology. Line 96 calls the high-order tensor contraction function \texttt{np.einsum} again. By looping over \texttt{Bs[i\_gp]} across the eight Gauss integration points, it recovers the microscopic stress fields of the solid elements under the six loading cases in parallel:
\begin{equation}
    \boldsymbol{\sigma}^{(I)}(\mathbf{q}_g) = \mathbf{C} \left[ \boldsymbol{\varepsilon}^{macro(I)}(z) - \mathbf{B}(\mathbf{q}_g) \mathbf{u}_e^{(I)} \right] \quad (I = 1, \dots, 6)
    \label{eq:stress_recovery}
\end{equation}
Here, the term $\mathbf{B}\mathbf{u}_e$ represents the relaxation effect of the microscopic fluctuation strain that arises as the structure re-establishes local equilibrium. The code directly evaluates the tensor difference \texttt{E\_macro - np.einsum(...)} and multiplies it by the constitutive matrix \texttt{C}, producing the full-field stress super-tensor \texttt{Sigma} with shape $(N_{active}, 6, 6)$.

After the microscopic 3D stress field has been obtained, it must be projected onto the macroscopic internal-resultant space of the 2D plate/shell. Lines 97-98 correspond to the zeroth- and first-order spatial moment integrals of the in-plane stress resultants in LPS-H. First, the code uses the array slice \texttt{Sigma[:, [0, 1, 5], :]} to extract the in-plane stress components from the 3D stress tensor, namely $\boldsymbol{\sigma}_p = [\sigma_{11}, \sigma_{22}, \sigma_{12}]^T$. It then uses \texttt{np.sum(..., axis=0)} to perform discrete summation and integration over the volume domain of the active elements.

In Line 97, the zeroth-order stress-resultant integral is extracted. This corresponds to the extraction of the membrane stiffness matrix $\mathbf{A}$ and the upper coupling block $\mathbf{B}$, where the program directly accumulates the in-plane stress over the volume and divides by the macroscopic projected area $S_{plate}$:

\begin{equation}
    \mathbf{ABD}_{[0:3, :]} = \frac{1}{S_{plate}} \sum_{e} \sum_{g=1}^{8} \boldsymbol{\sigma}_p^{(I)}(\mathbf{q}_g) |J| W_g
\end{equation}
    
In Line 98, the first-order stress-moment integral is extracted. This corresponds to the lower coupling block $\mathbf{B}^*$ and the bending stiffness matrix $\mathbf{D}$. The program multiplies the in-plane stress by the normal thickness coordinate \texttt{z\_active}, broadcast through \texttt{[:, None, None]}, to compute the first sectional moment, and again divides by the macroscopic area:

\begin{equation}
    \mathbf{ABD}_{[3:6, :]} = \frac{1}{S_{plate}} \sum_{e} \sum_{g=1}^{8} \boldsymbol{\sigma}_p^{(I)}(\mathbf{q}_g) \cdot z_e |J| W_g
\end{equation}

According to the Maxwell-Betti reciprocity theorem in linear elasticity, the $\mathbf{ABD}$ stiffness matrix derived from a conservative system is theoretically symmetric, namely $\mathbf{A}=\mathbf{A}^T$, $\mathbf{D}=\mathbf{D}^T$, and the upper and lower coupling blocks satisfy $\mathbf{B}=\mathbf{B}^{*T}$. In actual numerical computation, however, the directly extracted \texttt{ABD} matrix may exhibit slight asymmetry because of the iterative truncation tolerance of the GPU-based PCG solver and floating-point accumulation errors. To improve the rigor of the output and the stability of subsequent macroscopic finite element calls, the code performs the operation \texttt{(ABD + ABD.T) / 2.0} before returning the result.

%%%%%%%%%%%%%%%%%%%%%%%%%%
%%     Results Section  %%
%%%%%%%%%%%%%%%%%%%%%%%%%%
\section{Examples and Discussions}
\subsection{Base Case Execution}

\subsubsection{Primitive Sandwich Lattice Structure without Skins}
To demonstrate the usage logic and data flow of the open-source framework, a sheet-type Primitive structure is used below as the base case to illustrate the extraction procedure of the macroscopic effective $\mathbf{ABD}$ stiffness matrix. The relevant results can be reproduced by running the script \href{https://github.com/TopJournals/ThinWalledHomogenization/blob/main/examples/ex01_tpms_simulation.py}{\texttt{examples/ex01\_tpms\_simulation.py}} in the open-source repository.

First, the script defines the base-material properties and the geometric design parameters of the metamaterial plate. The base material is assigned Young's modulus $E = 1215$ MPa and Poisson's ratio $\nu = 0.35$. The physical thickness of the macroscopic plate/shell is set to $10$ mm, the mesh resolution of the unit cell is set to $N = 96$, and the target relative density is set to $0.15$, as shown in \textbf{Figure \ref{fig:model}a}. The program then calls the function \texttt{generate\_tpms\_voxel\_grid} from the implicit modeling module \href{https://github.com/TopJournals/ThinWalledHomogenization/blob/main/utils/tpms_generator.py}{\texttt{utils/tpms\_generator.py}} to discretize a 3D voxel matrix in space using an analytical level-set equation\citep{jiDesignOptimizationTPMSbased2025}:

\begin{figure*}[htbp]
    \centering
    \includegraphics[width=\textwidth]{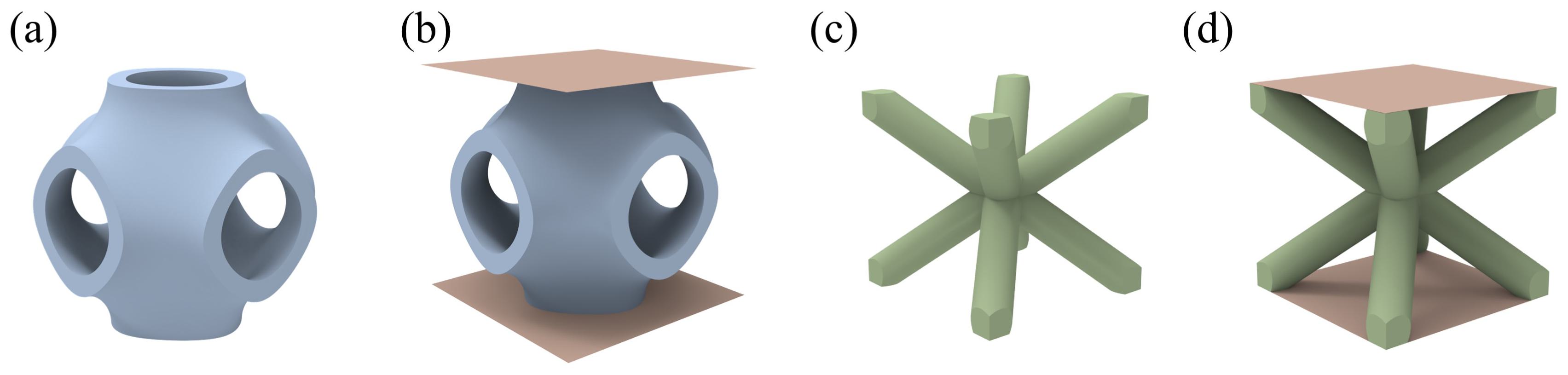}
    \caption{Schematic illustration of the four analysis cases. (a) TPMS sandwich structure without skins. (b) TPMS sandwich structure with skins. (c) BCC sandwich structure without skins. (d) BCC sandwich structure with skins.}
    \label{fig:model}
  \end{figure*}

\begin{lstlisting}[language=Python, numbers=left]
    # Step 2: Generate TPMS Voxel Model
    from utils.tpms_generator import generate_tpms_voxel_grid
    voxel_grid = generate_tpms_voxel_grid(
        tpms_type='Primitive',
        Nx=1, Ny=1, Nz=1,
        resolution=96,
        relative_density=0.15,  # Volume fraction
        is_sheet=True
    )
\end{lstlisting}

The generated \texttt{voxel\_grid} is then passed, together with the physical parameters, to the heterogeneous solver \texttt{homogenization\_plate} developed in Section 2.

\begin{lstlisting}[language=Python, numbers=left]
    # Step 3: Execute GPU-Accelerated LPS-H
    ABD_matrix = homogenization_plate(
        voxel=voxel_grid,
        E=E_base,
        nu=nu_base,
        thickness=thickness,
        Nx=Nx, Ny=Ny, Nz=Nz
    )
\end{lstlisting}
The entire solution takes 28.40 s and yields the following $6 \times 6$ macroscopic effective stiffness matrix for the equivalent continuum:

\begin{equation}
    \mathbf{ABD}_{Primitive} = 
    \begin{bmatrix}
        356.12 & 202.24 & 0.00 & \vdots & -0.09 & -0.01 & -0.04 \\
        202.24 & 356.12 & 0.00 & \vdots & -0.04 & -0.04 & -0.04 \\
        0.00 & 0.00 & 308.85 & \vdots & -0.04 & -0.04 & -0.06 \\
        \dots & \dots & \dots & \dots & \dots & \dots & \dots \\
        -0.09 & -0.04 & -0.04 & \vdots & 2229.51 & 1455.26 & 0.01 \\
        -0.01 & -0.04 & -0.04 & \vdots & 1455.26 & 2229.43 & 0.04 \\
        -0.04 & -0.04 & -0.06 & \vdots & 0.01 & 0.04 & 2024.48 
    \end{bmatrix}
\end{equation}

Quantitatively, for the in-plane membrane stiffness matrix $\mathbf{A}$, the diagonal entries $A_{11}$ and $A_{22}$ are nearly identical at 356.12 MPa$\cdot$mm, and the in-plane extension-shear coupling terms $A_{16}$ and $A_{26}$ are close to zero. This indicates that the Primitive structure is orthotropic in the macroscopic $x$-$y$ plane. For the extension-bending coupling matrix $\mathbf{B}$, most $B_{ij}$ values are on the order of $10^{-2}$ and can be regarded as numerical truncation error. Numerically, this indicates that the constructed sheet-type Primitive unit cell is approximately symmetric about the macroscopic geometric mid-surface, so in-plane deformation and out-of-plane bending are nearly decoupled. For the bending stiffness matrix $\mathbf{D}$, the values of $D_{11}$ and $D_{22}$ are also close, at about 2229.5 MPa$\cdot$mm$^3$, and the bending-twist coupling terms $D_{16}$ and $D_{26}$ are likewise close to zero. This suggests that parasitic geometric twisting is weak when the metamaterial plate is subjected to pure bending moments, which is characteristic of an orthotropic thin plate in bending.

\subsubsection{Primitive Sandwich Lattice Structure with Skins}

In practical engineering applications, Lattice-skin Plate Structures usually appear as sandwich structures composed of a lattice core and dense skins. Compared with a single lattice topology, the introduction of skins can protect the core cells from environmental exposure and improve the out-of-plane bending resistance of the plate/shell. The following example shows a typical workflow for analyzing a Lattice-skin Plate Structure with this framework. By adding a \texttt{NumPy} padding operation after the voxel-generation block of \href{https://github.com/TopJournals/ThinWalledHomogenization/blob/main/examples/ex01_tpms_simulation.py}{\path{ex01_tpms_simulation.py}}, a skinned model can be created:

\begin{lstlisting}[language=Python]
    voxel_grid = np.pad(
        voxel_grid,
        pad_width=((0, 0), (0, 0), (2, 2)),
        mode='constant',
        constant_values=1
    )
\end{lstlisting}

For the sheet-type Primitive structure with added skins, the relative density and thickness are kept at 0.15 and 10 mm, respectively. The model is shown in \textbf{Figure \ref{fig:model}b}. The program outputs the following effective $\mathbf{ABD}$ stiffness matrix:

\begin{equation}
    \mathbf{ABD}_{Skinned} = 
    \begin{bmatrix}
        973.08 & 412.50 & 0.00 & \vdots & -0.08 & -0.06 & -0.03 \\
        412.50 & 973.08 & 0.00 & \vdots & -0.06 & -0.08 & -0.03 \\
        0.00 & 0.00 & 512.03 & \vdots & -0.04 & -0.04 & -0.07 \\
        \dots & \dots & \dots & \dots & \dots & \dots & \dots \\
        -0.08 & -0.06 & -0.04 & \vdots & 16538.85 & 6436.16 & 0.03 \\
        -0.06 & -0.08 & -0.04 & \vdots & 6436.16 & 16538.85 & 0.03 \\
        -0.03 & -0.03 & -0.07 & \vdots & 0.03 & 0.03 & 6543.43
    \end{bmatrix}
\end{equation}

The results give $A_{11} = A_{22} = 973.08$ MPa$\cdot$mm and $D_{11} = D_{22} = 16538.85$ MPa$\cdot$mm$^3$. The addition of skins increases the in-plane membrane stiffness by about 170\%, while the bending stiffness $D_{11}$ rises from about 2229 to 16538, an increase of about 640\%. This indicates that Lattice-skin Plate Structures can provide higher stiffness in applications, such as aerospace, where bending resistance is important.

\subsubsection{BCC Sandwich Lattice Structure without Skins}

To further verify the generality and numerical stability of the open-source framework for truss-based structures, a classical body-centered cubic (BCC) lattice is selected as a validation case. The results can be reproduced by running the script \href{https://github.com/TopJournals/ThinWalledHomogenization/blob/main/examples/ex01_2_lattice_simulation.py}{\path{examples/ex01_2_lattice_simulation.py}} in the open-source repository. The same physical parameters as in the TPMS case are used: $E = 1215$ MPa, $\nu = 0.35$, thickness 10 mm, and target relative density 0.15. The model is shown in \textbf{Figure \ref{fig:model}c}. The effective $\mathbf{ABD}$ stiffness matrix of the structure without skins is then obtained through the implicit truss modeling module \href{https://github.com/TopJournals/ThinWalledHomogenization/blob/main/utils/lattice_generator.py}{\path{utils/lattice_generator.py}}:

\begin{equation}
    \mathbf{ABD}_{BCC} = 
    \begin{bmatrix}
        68.84 & 32.12 & -0.07 & \vdots & 0.21 & -0.19 & 0.18 \\
        32.12 & 68.84 & -0.01 & \vdots & -0.16 & 0.18 & 0.19 \\
        -0.07 & -0.01 & 256.47 & \vdots & 0.20 & 0.16 & 0.82 \\
        \dots & \dots & \dots & \dots & \dots & \dots & \dots \\
        0.21 & -0.16 & 0.20 & \vdots & 625.52 & 234.58 & -0.31 \\
        -0.19 & 0.18 & 0.16 & \vdots & 234.58 & 625.55 & 0.17 \\
        0.18 & 0.19 & 0.82 & \vdots & -0.31 & 0.17 & 1674.86
    \end{bmatrix}
\end{equation}

This matrix shows that the BCC lattice structure also exhibits orthotropic behavior in its macroscopic mechanical response. Its in-plane membrane stiffnesses $A_{11}$ and $A_{22}$ are both 68.84 MPa$\cdot$mm, and the extension-shear coupling terms are close to zero. Compared with the pure tensile stiffness, the in-plane shear stiffness $A_{66}$ of the BCC structure is 256.47 MPa$\cdot$mm, which is larger than $A_{11}$. This is closely related to its topology of spatial diagonal struts: the BCC topology is bending-dominated under principal-axis tension/compression, but becomes stretch-dominated under shear. In addition, the entries of the extension-bending coupling block $\mathbf{B}$ remain within a small numerical truncation range, indicating macroscopic geometric mid-surface symmetry. For bending resistance, $D_{11}$ and $D_{22}$ are close to each other, at about 625.5 MPa$\cdot$mm$^3$, and the bending-twist coupling terms also approach zero, which is consistent with the bending behavior of an orthotropic thin plate.

\subsubsection{BCC Sandwich Lattice Structure with Skins}

Similarly, to investigate the mechanical enhancement effect of dense skins on a discrete lattice architecture, top and bottom \texttt{NumPy} padding layers are added to the BCC unit-cell model above, and the homogenization analysis is repeated. The model is shown in \textbf{Figure \ref{fig:model}d}. The resulting macroscopic effective $\mathbf{ABD}$ stiffness matrix of the skinned BCC sandwich plate is

\begin{equation}
    \mathbf{ABD}_{BCC\_Skinned} = 
    \begin{bmatrix}
        667.89 & 241.00 & -0.08 & \vdots & 1.20 & 0.18 & 0.18 \\
        241.00 & 667.78 & -0.05 & \vdots & 0.22 & 0.89 & 0.23 \\
        -0.08 & -0.05 & 437.42 & \vdots & 0.19 & 0.19 & 0.80 \\
        \dots & \dots & \dots & \dots & \dots & \dots & \dots \\
        1.20 & 0.22 & 0.19 & \vdots & 14513.15 & 5072.29 & -0.41 \\
        0.18 & 0.89 & 0.19 & \vdots & 5072.29 & 14511.40 & -0.26 \\
        0.18 & 0.23 & 0.80 & \vdots & -0.41 & -0.26 & 5921.17
    \end{bmatrix}
\end{equation}

Compared with the pure BCC lattice structure without skins, the introduction of skins increases the overall effective stiffness. The in-plane principal stiffness $A_{11}$ increases from 68.84 MPa$\cdot$mm to 667.89 MPa$\cdot$mm, an increase of about 870\%. The bending stiffness $D_{11}$ increases from 625.52 MPa$\cdot$mm$^3$ to 14513.15 MPa$\cdot$mm$^3$, an increase of about 2200\%. This indicates that skins located far from the neutral surface improve the bending efficiency of the section and dominate when the thin-walled plate/shell carries out-of-plane bending moments. Together with the TPMS examples above, these results show that the framework has good applicability and stability in cross-dimensional stiffness tensor assembly. They also indicate that LPS-H can be used for metamaterial laminated structures with multiscale geometric features, providing an analysis tool for the development and design of multifunctional lightweight Lattice-skin Plate Structures.

\subsection{Mesh Convergence Analysis}
\begin{figure*}[htbp]
  \centering
  \includegraphics[width=0.8\textwidth]{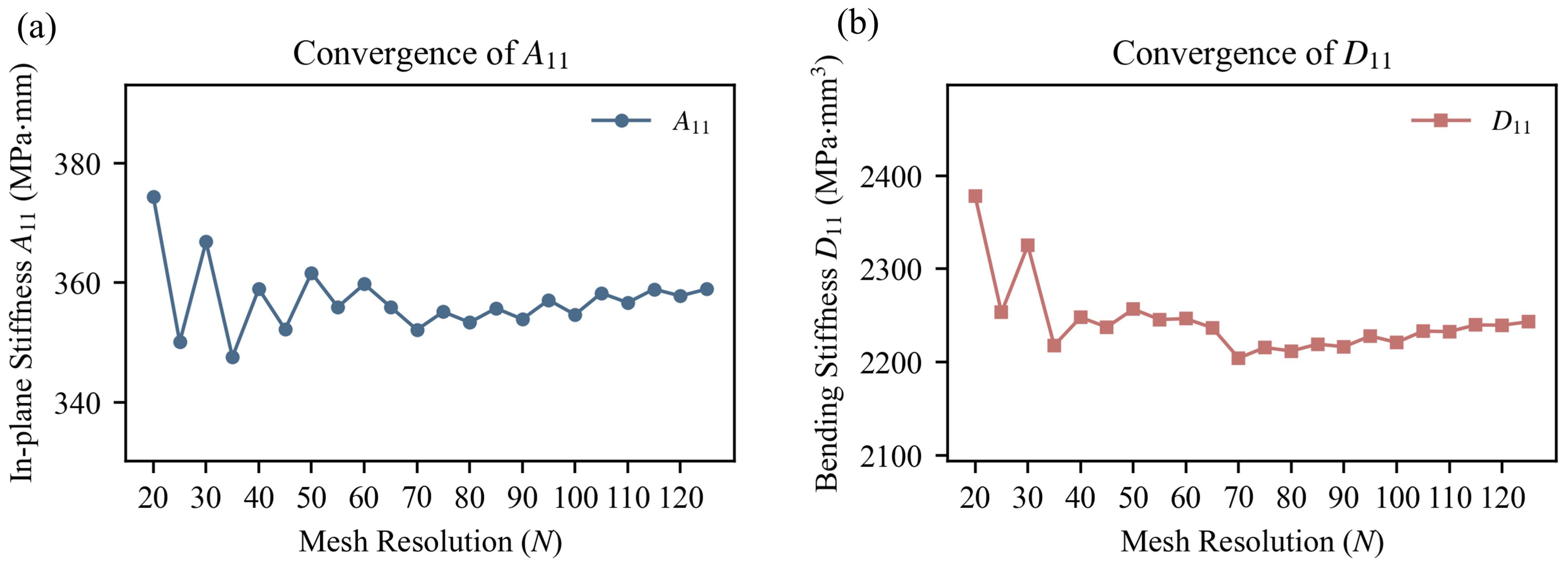}
  \caption{Mesh convergence analysis of the sheet-type Primitive structure. (a) Evolution of the in-plane membrane stiffness $A_{11}$ with respect to the resolution $N$. (b) Evolution of the bending stiffness $D_{11}$ with respect to the resolution $N$.}
  \label{fig:convergence}
\end{figure*}

In voxel-based finite element homogenization, the mesh resolution affects both the geometric fidelity of the implicit surface boundary and the number of DOFs in the discrete model. To balance computational accuracy and efficiency, a mesh convergence analysis is carried out for the LPS-H framework. The sheet-type Primitive structure is selected as the reference case, while the cell array and the relative density of 0.15 are kept unchanged. The one-dimensional mesh resolution of the unit cell, $N$, is increased from 20 to 125 with a sampling interval of 5. The in-plane principal tensile stiffness $A_{11}$ and the out-of-plane principal bending stiffness $D_{11}$ are extracted from the effective stiffness matrix at each resolution. Their convergence histories are shown in \textbf{Figure \ref{fig:convergence}}.

The results show that when the one-dimensional mesh resolution is low, the predicted in-plane and out-of-plane stiffness values exhibit noticeable fluctuations. This is related to the stair-step modeling effect produced when a coarse voxel grid approximates a complex TPMS surface. As the mesh resolution increases, the voxel model approximates the implicit surface more accurately and the stair-step effect gradually weakens. When $N$ exceeds 60, the values of $A_{11}$ and $D_{11}$ become stable and show asymptotic convergence.

Further increasing the resolution can improve the approximation of the analytical geometry, but the total number of DOFs in 3D voxel-based finite elements is proportional to $N^3$, so the memory cost and solution time also increase. Based on the convergence analysis and the numerical stability discussion above, a unit-cell mesh resolution in the range $N = 60 \sim 65$ is a reasonable choice for LPS-H performance evaluation of thin-walled TPMS plate/shells. This range balances mechanical accuracy and GPU-based solution efficiency, and can be used as a practical compromise in engineering analysis.

\subsection{Anisotropy and Extension-Twist Coupling in TPMS Plates}
A benchmark comparison between LVS-H and LPS-H is presented below. The analysis considers four representative sheet-type TPMS structures, namely Gyroid, Diamond, Primitive, and I-WP. Both homogenization methods use a mesh resolution of $128^3$, a relative density of 0.15, and a physical thickness of $10$ mm. Only a single truncated period is retained through the thickness, i.e., $N_z = 1$. Heat maps comparing the effective $\mathbf{ABD}$ stiffness matrices extracted by LPS-H with the predictions from LVS-H, together with the corresponding relative-error distributions, are shown in \textbf{Figure \ref{fig:fig_r1sheet}}. The results indicate that LVS-H exhibits bias in both stiffness prediction and coupling-response characterization for finite-thickness structures.

\begin{figure*}[htbp]
  \centering
  \includegraphics[width=\textwidth]{fig_r1sheet.jpg}
  \caption{Heat-map comparison of effective $\mathbf{ABD}$ stiffness predictions from LVS-H and LPS-H for sheet-type TPMS structures, together with the corresponding relative-error distributions.}
  \label{fig:fig_r1sheet}
\end{figure*}

Because LVS-H imposes 3D-PBC in all three spatial directions, the equivalent model tends to preserve the symmetry of the spatial lattice. As a result, LVS-H predicts identical in-plane tensile stiffness for the Gyroid structure, namely $A_{11} = A_{22} = 674.00$ MPa$\cdot$mm. When the continuous Gyroid surface is physically truncated into a single-layer thin plate with thickness $10$ mm, however, the asymmetric truncation of the upper and lower surfaces breaks the original body-centered cubic symmetry and induces macroscopic in-plane orthotropy. The LPS-H results show a clear stiffness difference, with $A_{11} = 448.68$ MPa$\cdot$mm and $A_{22} = 627.66$ MPa$\cdot$mm. Relative to the LPS-H results, the LVS-H predictions of tensile stiffness are higher by $50.22\%$ and $7.49\%$ in the $x$ and $y$ directions, respectively.

LVS-H also overestimates the out-of-plane bending stiffness matrix $\mathbf{D}$. The bending-stiffness prediction errors for the four structures are 151.09\%, 15.35\%, 221.11\%, and 37.24\%, respectively. This indicates that LVS-H mathematically assumes a uniform material distribution through the thickness direction $z$, which changes the sectional moment of inertia of the structure.

In addition, the three-dimensional periodic boundaries used in LVS-H cannot fully reflect the intrinsic asymmetric coupling deformation mechanisms of the topology. Taking the sheet-type Diamond structure as an example, the LPS-H results show nonzero terms $B_{16}$ and $B_{26}$ in the extension-bending coupling matrix, with values reaching $285.07$ MPa$\cdot$mm$^2$. These nonzero terms indicate a macroscopic extension-twist coupling effect. Under in-plane uniaxial tension, the plate/shell may therefore undergo macroscopic out-of-plane twisting deformation.

\subsection{Thickness Size Effect and Asymptotic Convergence}

\begin{figure*}[htbp]
  \centering
  \includegraphics[width=\textwidth]{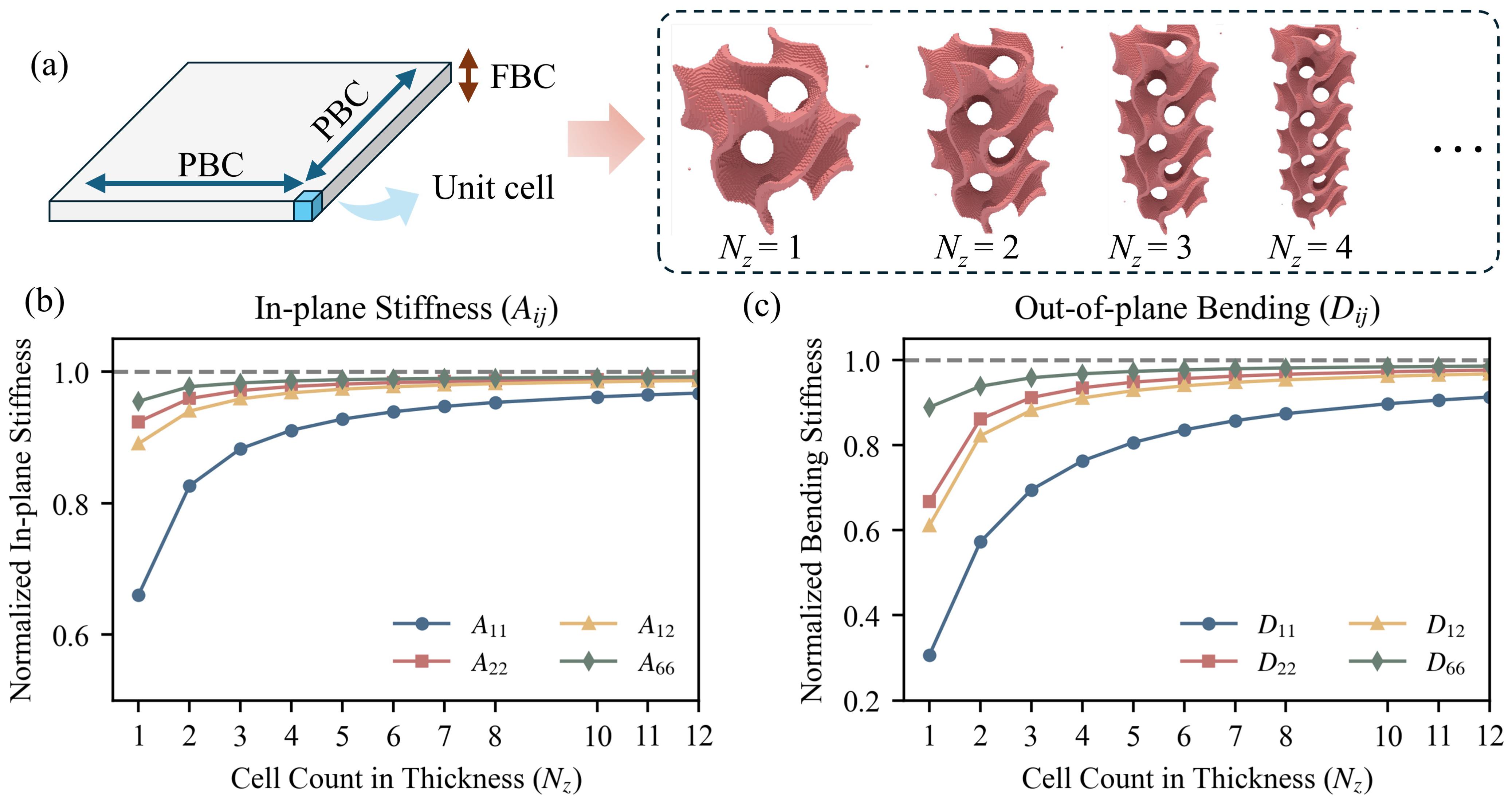}
  \caption{Thickness size-effect analysis of the sheet-type Gyroid structure. (a) Schematic of the geometric evolution as the number of layers through the thickness, $N_z$, increases; in-plane periodic boundary conditions (PBC) are applied and out-of-plane free boundary conditions (FBC) are retained. (b) Convergence curves of the normalized in-plane membrane stiffness $A_{ij}$ versus $N_z$. (c) Convergence curves of the normalized bending stiffness $D_{ij}$ versus $N_z$.}
  \label{fig:size_effect}
\end{figure*}

In classical continuum mechanics, as the plate/shell thickness increases, the influence of surface boundary conditions on macroscopic stiffness gradually decays, and finite-thickness plate theory asymptotically approaches 3D solid theory. To evaluate the physical consistency and generality of the open-source code, the thickness size effect of metamaterial plates is analyzed below. The results can be reproduced by running \href{https://github.com/TopJournals/ThinWalledHomogenization/blob/main/examples/ex02_multiple_cells.py}{\path{examples/ex02_multiple_cells.py}}. Taking the sheet-type Gyroid structure as an example, the in-plane cell count is kept at $1 \times 1$, while the number of cells through the thickness, $N_z$, is increased from 1 to 12, as shown in \textbf{Figure \ref{fig:size_effect}(a)}. For convenience of comparison, each stiffness component computed by LPS-H is normalized by the theoretical stiffness value derived from LVS-H together with the corresponding formulas. The convergence curves of the normalized in-plane membrane stiffness $A_{ij}$ and out-of-plane bending stiffness $D_{ij}$ versus the layer number $N_z$ are shown in \textbf{Figure \ref{fig:size_effect}(b)} and \textbf{(c)}, respectively. The dashed line at 1.0 represents the LVS-H prediction benchmark.

The results show that when the number of layers is small, the predictions of LPS-H and LVS-H differ substantially. For the thin plate with $N_z = 1$, the normalized in-plane principal stiffnesses $A_{11}$ and $A_{22}$ are 0.66 and 0.92, respectively. Relative to the LPS-H results, the corresponding LVS-H predictions are higher by 51.5\% and 8.7\%. The discrepancy is even more pronounced for the bending stiffness, where the normalized values of $D_{11}$ and $D_{22}$ are 0.31 and 0.67, corresponding to deviations of 226.3\% and 49.7\%. As the number of layers through the thickness increases, the normalized stiffness curves increase monotonically and gradually approach 1.0. For the in-plane stiffness, the normalized values exceed 0.90 and 0.95 at 4 and 8 layers, respectively. For the out-of-plane stiffness, the normalized values exceed 0.90 when the number of layers reaches 11.

\section{Extensions}

\subsection{Extension to Multimaterial Lattice-skin Plate Structures}

With the development of additive manufacturing, metamaterial plate/shells printed from multiple base materials are becoming a new paradigm in structural design. In multimaterial systems, the constitutive relation is no longer a global constant. Instead, it becomes a nonuniform tensor field $\mathbf{C}(\mathbf{x})$ that varies with spatial position. The following shows how the single-phase LPS-H framework can be extended to multimaterial homogenization with only modest code modifications.

From both the physical and mathematical viewpoints, the main change introduced by multiple materials is that each discrete voxel element $e$ has its own local material properties, such as Young's modulus $E_e$ and Poisson's ratio $\nu_e$. Accordingly, the integral formula for the element stiffness matrix must be modified to depend on the local constitutive tensor $\mathbf{C}_e$:

\begin{equation}
    \mathbf{K}_e^{(e)} = \int_{V_e} \mathbf{B}^T \mathbf{C}_e \mathbf{B} dV = \sum_{g=1}^{8} \mathbf{B}^T(\mathbf{q}_g) \mathbf{C}_e \mathbf{B}(\mathbf{q}_g) |J| W_g
\end{equation}

In the Python implementation, preserving the loop-free computational efficiency of the framework requires abandoning the traditional element-by-element traversal used in finite element codes and continuing to exploit the advantages of \texttt{NumPy} multidimensional tensor operations. The main modifications are concentrated in the following three modules:

\subsubsection*{Initialization of the material constitutive tensor field}
\begin{figure*}[htbp]
  \centering
  \includegraphics[width=\textwidth]{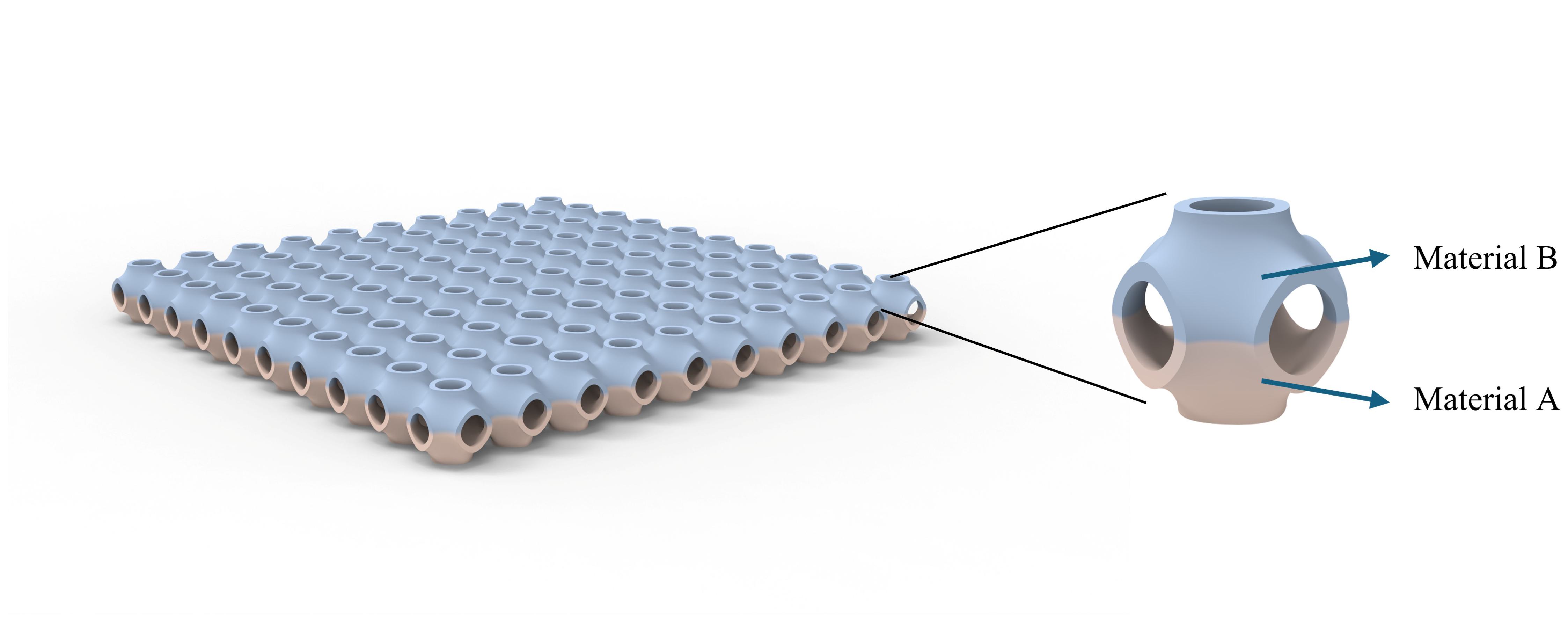}
  \caption{Schematic of a bimaterial laminated metamaterial plate/shell.}
  \label{fig:multimaterial}
\end{figure*}

In the original code, \texttt{C} is a two-dimensional array with shape $6 \times 6$. For multimaterial analysis, a three-dimensional super-tensor \texttt{C\_active} with shape \texttt{($N_{active}$, 6, 6)} must be generated for the active elements according to the spatial position or phase state of the voxels. In practical metamaterial engineering, two multimaterial-definition scenarios are common:

\textbf{Scenario A: regular partition based on spatial coordinates}\\
Consider a typical bimaterial laminated plate, as shown in \textbf{Figure \ref{fig:multimaterial}}, in which the base material has Young's modulus $500$ MPa in the upper half through the thickness, $z > 0$, and $1215$ MPa in the lower half, $z \le 0$. A vectorized initialization can be performed using the conditional-selection functionality of \texttt{NumPy} based on the thickness coordinates \texttt{z\_active} of the elements.

\textbf{Scenario B: arbitrary irregular phase mapping based on a voxel matrix}\\
More generally, the material distribution inside many multimaterial structures is irregular. In this case, an external material-label matrix \texttt{material\_grid}, consistent with the structural geometry and of size $N_x \times N_y \times N_z$, is usually maintained. Here, the value $1$ represents the first material, $2$ represents the second material, and so on. In the code, Boolean masking can be used to extract the material IDs of the active elements, and \texttt{np.select} can then realize loop-free mapping of multiphase material properties.

Efficient vectorized implementations of these two scenarios in Python are shown below:

\begin{lstlisting}[language=Python]
    # ====================================================================
    # Scenario A: Based on spatial coordinates (e.g., Bi-material plate)
    # ====================================================================
    E_active = np.where(z_active > 0, 500.0, 1215.0) 
    
    # ====================================================================
    # Scenario B: Based on an explicit material matrix (e.g., Custom field)
    # ====================================================================
    # Assume `material_grid` is passed into the function alongside `voxel`
    # Extract material IDs exclusively for the solid active elements
    # mat_id_active = material_grid[active_mask]  
    
    # Map material IDs to real physical parameters without any loops
    conditions = [mat_id_active == 1, mat_id_active == 2]
    choices = [1215.0, 500.0]  # E.g., Material 1 is 1215 MPa, Material 2 is 500 MPa
    
    # Assign default modulus (e.g., 10.0 MPa) to any unmapped IDs as a fallback
    E_active = np.select(conditions, choices, default=10.0)
    
    # Finally, build the heterogenous constitutive super-tensor
    # `get_multi_material_elasticity` handles the array of E efficiently
    C_active = get_multi_material_elasticity(E_active, nu_base) # Shape: (N_active, 6, 6)
\end{lstlisting}

At the same time, a new function, \texttt{get\_multi\_material\_elasticity}, is introduced to replace the original \texttt{get\_isotropic\_elasticity} function.

\begin{lstlisting}[language=Python]
def get_multi_material_elasticity(E_array, nu=0.3):
    lam = E_array * nu / ((1 + nu) * (1 - 2 * nu))
    mu = E_array / (2 * (1 + nu))
    N = len(E_array)
    C_active = np.zeros((N, 6, 6))
    C_active[:, 0, 1] = C_active[:, 0, 2] = C_active[:, 1, 0] = lam
    C_active[:, 1, 2] = C_active[:, 2, 0] = C_active[:, 2, 1] = lam
    C_active[:, 0, 0] = C_active[:, 1, 1] = C_active[:, 2, 2] = lam + 2 * mu
    C_active[:, 3, 3] = C_active[:, 4, 4] = C_active[:, 5, 5] = mu
    return C_active
\end{lstlisting}

\subsubsection*{Vectorized assembly of heterogeneous element stiffness and loads}
Because the constitutive tensor $\mathbf{C}_e$ differs from one element to another, the original operation \texttt{np.tile}, which replicates a single \texttt{Ke}, is no longer applicable. Instead, higher-order tensor contraction must be used to compute the individual stiffness matrices of all active elements simultaneously. The original assembly logic can be replaced by the following \texttt{np.einsum}-based operations:

\begin{lstlisting}[language=Python]
    # Compute active element stiffness matrices simultaneously
    Ke_active = np.zeros((len(edofMat), 24, 24))
    for i in range(8):
        # np.einsum notation: 'ji' (B.T), 'ekj' (C_active), 'kl' (B) -> 'eil' (Ke for each element e)
        Ke_active += np.einsum('ji,ekj,kl->eil', Bs[i], C_active, Bs[i]) * detJ
    
    # Flatten the heterogeneous stiffness array directly for COO matrix
    sK = Ke_active.flatten() 
\end{lstlisting}

For the calculation of the initial macroscopic loads, the original broadcast multiplication must likewise be upgraded to a contraction that accounts for the element dimension:

\begin{lstlisting}[language=Python]
    # Modify local load calculation for multi-material
    F_ele = sum([np.einsum('ji,ejl->eil', Bs[i], 
                 np.einsum('eij,ejl->eil', C_active, E_macro)) * detJ for i in range(8)])
\end{lstlisting}
Here, \texttt{eij,ejl->eil} specifies one-to-one multiplication of tensors along the first dimension \texttt{e}, where \texttt{e} is the element index, thereby realizing the physical mapping of material heterogeneity.

\subsubsection*{Heterogeneous stress recovery}
Similarly, during the final macroscopic area-integration stage for extracting the $\mathbf{ABD}$ matrix, recovery of the physical stress \texttt{Sigma} must also depend on the local \texttt{C\_active}. The stress calculation in Line 96 of the original code should therefore be modified as follows:

\begin{lstlisting}[language=Python]
    # Recover true physical stress utilizing local constitutive tensors
    Sigma = np.einsum('eij,ejl->eil', C_active, 
                      E_macro - np.einsum('ij,ejl->eil', Bs[i_gp], U[edofMat, :]))
\end{lstlisting}

By introducing the dimension \texttt{'e'} into the \texttt{einsum} string specification above, LPS-H analysis can be extended to multimaterial, multiphase, and even damage-containing materials defined by arbitrary voxels, without increasing the nesting depth of Python \texttt{for} loops.

Running the script \href{https://github.com/TopJournals/ThinWalledHomogenization/blob/main/examples/ex03_multimaterial_simulation.py}{\path{examples/ex03_multimaterial_simulation.py}} in the open-source repository produces the following $6 \times 6$ macroscopic effective stiffness matrix for the bimaterial case:

\begin{equation}
    \mathbf{ABD}_{Bi-material} = 
    \begin{bmatrix}
        245.86 & 147.49 & 0.00 & \vdots & -229.85 & -176.77 & -0.02 \\
        147.49 & 245.86 & 0.00 & \vdots & -176.77 & -229.85 & -0.02 \\
        0.00 & 0.00 & 217.04 & \vdots & -0.02 & -0.02 & -223.64 \\
        \dots & \dots & \dots & \dots & \dots & \dots & \dots \\
        -229.85 & -176.77 & -0.02 & \vdots & 1564.58 & 1033.34 & 0.03 \\
        -176.77 & -229.85 & -0.02 & \vdots & 1033.34 & 1564.58 & 0.03 \\
        -0.02 & -0.02 & -223.64 & \vdots & 0.03 & 0.03 & 1417.09 
    \end{bmatrix}
\end{equation}

Compared with the $\mathbf{ABD}$ matrix of the single-phase homogeneous material in Section 4, the $\mathbf{B}$ matrix entries of the multimaterial structure are no longer zero. The main terms, such as $B_{11}$ and $B_{12}$, are $-229.85$ and $-176.77$ MPa$\cdot$mm$^2$, respectively. This indicates that when pure in-plane tension is applied to the bimaterial metamaterial plate/shell, macroscopic bending deformation may occur because of stress mismatch between the upper and lower layers.

\subsection{Extension to Thermal Conduction Homogenization}

With the continuing miniaturization and increasing power density of aerospace electronic devices, metamaterial plate/shells capable of integrating structural and thermal functions have become of interest for thermal protection and heat-dissipation design. In continuum mechanics, the weak forms of the linear elastic static equations and the steady-state heat conduction equations are highly isomorphic. The following shows how this similarity can be used to extend the LPS-H mechanical code into a tool for predicting effective in-plane thermal conductivity.

In steady-state heat conduction theory, the basic mapping of physical variables is as follows: the displacement vector $\mathbf{u}$ is replaced by the temperature field $T$; the strain tensor $\boldsymbol{\varepsilon}$ maps to the temperature gradient $\nabla T$; the stress tensor $\boldsymbol{\sigma}$ maps to the heat-flux vector $\mathbf{q}$; and the elastic constitutive matrix $\mathbf{C}$ corresponds to the thermal conductivity tensor $\mathbf{k}$. Fourier's law of heat conduction can be written as $\mathbf{q} = -\mathbf{k} \nabla T$.

\subsubsection*{Scalar reduction of the global DOF field}
In the mechanical analysis, each node has three translational DOFs, namely $(u, v, w)$. In the heat-conduction problem, however, each node has only one scalar DOF, namely the temperature $T$. Therefore, the tensorized DOF mapping function \texttt{build\_tensor\_dof\_mapping} must be modified for the global DOF tensor \texttt{dof\_tensor}:

\begin{lstlisting}[language=Python]
    # Thermal DOF mapping: 1 Degree of Freedom (Temperature) per node
    node_indices = np.arange((nx + 1) * (ny + 1) * (nz + 1)).reshape(nx + 1, ny + 1, nz + 1)
    dof_tensor = node_indices.copy() # No need for the last axis of length 3
    
    # 2D-PBC bounds apply exactly as before
    dof_tensor[nx, :, :] = dof_tensor[0, :, :]
    dof_tensor[:, ny, :] = dof_tensor[:, 0, :]
\end{lstlisting}

\subsubsection*{Vectorized construction of the element thermal conduction matrix and loads}
For an isotropic base material, the thermal conductivity tensor is the diagonal matrix $\mathbf{k} = k_{base} \mathbf{I}_{3 \times 3}$. The element-level conduction matrix becomes
\begin{equation}
    \mathbf{K}_T^{(e)} = \int_{V_e} (\nabla \mathbf{N})^T \mathbf{k} (\nabla \mathbf{N}) dV
\end{equation}
Because the temperature field is a scalar field, the original strain-displacement matrix $\mathbf{B}$ of size $6 \times 24$ must be replaced by the $3 \times 8$ shape-function gradient matrix $\nabla \mathbf{N}$. In the original code, the derivative assembly is therefore simplified to the matrix \texttt{dN}:

\begin{lstlisting}[language=Python]
    # Element Thermal Conductivity Matrix
    Kt_e, gradNs = np.zeros((8, 8)), []
    for q in nodes / np.sqrt(3):
        # dN is the 3x8 gradient matrix of shape functions
        dN = 0.125 * np.array([(1 + q[1] * nodes[:, 1]) * (1 + q[2] * nodes[:, 2]) * nodes[:, 0] * (2 / dx), (1 + q[0] * nodes[:, 0]) * (1 + q[2] * nodes[:, 2]) * nodes[:, 1] * (2 / dy), (1 + q[0] * nodes[:, 0]) * (1 + q[1] * nodes[:, 1]) * nodes[:, 2] * (2 / dz)])
        gradNs.append(dN)
        Kt_e += dN.T @ k_tensor @ dN * detJ
\end{lstlisting}

\subsubsection*{Macroscopic temperature-gradient loading and extraction of effective thermal conductivity}
For the 2D dimension-reduced heat-transfer analysis of a sheet-type plate/shell, the quantity of interest is the in-plane heat-dissipation efficiency. Therefore, only two macroscopic unit temperature-gradient cases need to be applied, namely the temperature gradients along the $x$ and $y$ axes, $\nabla T_{macro} = [1, 0, 0]^T$ and $[0, 1, 0]^T$. The macroscopic excitation tensor \texttt{Grad\_macro} thus becomes a matrix with shape \texttt{(3, 2)}.

After the same PCG-based heterogeneous solver is used to obtain the microscopic temperature fluctuation field $\mathbf{T}_{fluc}$, the macroscopic effective thermal conductivity matrix $\mathbf{k}^{hom}$ can be extracted by integrating the in-plane heat-flux density. Here, $\mathbf{T}_{fluc}$ has shape $N_{dofs} \times 2$, while $\mathbf{k}^{hom}$ has size $2 \times 2$:

\begin{lstlisting}[language=Python]
    k_hom = np.zeros((2, 2))
    for i_gp in range(8):
        # Recover True Heat Flux (q = k * (Grad_macro + gradN * T_fluc))
        Flux = np.einsum('ij, jl->il', k_tensor, Grad_macro - np.einsum('ij, ejl->eil', gradNs[i_gp], T_fluc[edofMat, :]))
        
        # Extract 2x2 in-plane thermal conductivity matrix via area integration
        k_hom += np.sum(Flux[:, 0:2, :], axis=0) * detJ / plate_area 
\end{lstlisting}

Running the script \href{https://github.com/TopJournals/ThinWalledHomogenization/blob/main/examples/ex04_tpms_thermal_simulation.py}{\path{examples/ex04_tpms_thermal_simulation.py}} in the open-source repository produces the following $2 \times 2$ macroscopic effective in-plane thermal conductivity matrix:

\begin{equation}
    \mathbf{k}^{hom} = 
    \begin{bmatrix}
        60.19 & 0.00 \\
        0.00 & 60.19
    \end{bmatrix}
\end{equation}

The above calculation results are based on the following standard test parameters: the analysis object is the sheet-type Primitive structure, the unit-cell relative density is $\bar{\rho} = 0.15$, the overall plate thickness is $h = 10\text{ mm}$, and the homogeneous base material is assigned a thermal conductivity of $k_s = 60.5\text{ W/(m}\cdot\text{K)}$.

\section{Conclusion}
For the dimensional-reduction analysis of Lattice-skin Plate Structures, the open-source Python implementation of LPS-H alleviates the limitations of LVS-H in the analysis of finite-thickness thin-walled structures. Because LVS-H is constrained by the periodic assumption through the thickness, it may overestimate out-of-plane bending stiffness and neglect the extension-bending and extension-twist coupling effects caused by surface truncation when finite-thickness thin-walled structures are analyzed. The proposed computational framework introduces mixed boundary conditions that are periodic in plane and free out of plane. These boundary conditions help reflect the underlying mechanical mechanisms and enable reduced-order extraction of the macroscopic effective $\mathbf{ABD}$ stiffness matrix.

At the algorithmic level, the program replaces the conventional element-by-element nested assembly loops in finite element analysis with a loop-free topological DOF mapping algorithm based on multidimensional array slicing, together with high-order tensor contraction using \texttt{np.einsum}. These techniques enable vectorized assembly of the global stiffness and load matrices. Combined with the non-blocking CUDA-stream concurrency mechanism provided by \texttt{CuPy} and the Jacobi-preconditioned conjugate gradient solver, the heterogeneous acceleration framework can solve high-fidelity metamaterial unit-cell models containing millions of DOFs.

The benchmark tests and convergence analyses show that complex metamaterial plate/shells exhibit thickness size effects and macroscopic anisotropy. The results further verify the numerical stability and theoretical consistency of the algorithm across different mesh resolutions. Owing to its generalized vectorized programming architecture, the homogenization code can be extended to the analysis of multiphase laminated structures and to multiphysics performance evaluation such as steady-state heat conduction. The open-source code provided here can serve as a foundational analysis tool for computational mechanics education and for the high-fidelity structural design and topology optimization of multifunctional thin-walled metamaterials.

\backmatter
\begin{appendices}

\section{A 99-line Python Code for ABD Matrix Homogenization}\label{sec:appendix_code}

\lstset{language=Python, basicstyle=\ttfamily\small, breaklines=true}
\begin{lstlisting}
import numpy as np
from scipy.sparse import coo_matrix
import cupy as cp
import cupyx.scipy.sparse as cpsp
import cupyx.scipy.sparse.linalg as cpspla


def get_isotropic_elasticity(E=1.0, nu=0.3):
    lam, mu = E * nu / ((1 + nu) * (1 - 2 * nu)), E / (2 * (1 + nu))  # Calculate Lame parameters
    C = np.zeros((6, 6))  # Initialize 6x6 constitutive matrix
    C[0:3, 0:3] = lam
    np.fill_diagonal(C[0:3, 0:3], lam + 2 * mu)
    np.fill_diagonal(C[3:6, 3:6], mu)
    return C


def compute_element_stiffness(C, dx, dy, dz):
    Ke, Bs = np.zeros((24, 24)), []
    nodes = np.array([[-1, -1, -1], [1, -1, -1], [1, 1, -1], [-1, 1, -1], [-1, -1, 1], [1, -1, 1], [1, 1, 1], [-1, 1, 1]])  # 8-node local coordinates
    detJ = dx * dy * dz / 8.0
    for q in nodes / np.sqrt(3):  # Loop over 8 Gauss integration points
        dN = 0.125 * np.array([(1 + q[1] * nodes[:, 1]) * (1 + q[2] * nodes[:, 2]) * nodes[:, 0] * (2 / dx), (1 + q[0] * nodes[:, 0]) * (1 + q[2] * nodes[:, 2]) * nodes[:, 1] * (2 / dy), (1 + q[0] * nodes[:, 0]) * (1 + q[1] * nodes[:, 1]) * nodes[:, 2] * (2 / dz)])  # Cartesian derivatives
        B = np.zeros((6, 24))  # Strain-displacement matrix
        B[0, 0::3], B[1, 1::3], B[2, 2::3] = dN[0], dN[1], dN[2]
        B[3, 1::3], B[3, 2::3] = dN[2], dN[1]
        B[4, 0::3], B[4, 2::3] = dN[2], dN[0]
        B[5, 0::3], B[5, 1::3] = dN[1], dN[0]
        Bs.append(B)  # Store B matrices for stress recovery
        Ke += B.T @ C @ B * detJ  # Accumulate local stiffness
    return Ke, Bs, detJ


def build_tensor_dof_mapping(voxel, dx, dy, dz, thickness):
    nx, ny, nz = voxel.shape  # Enforce XYZ indexing convention
    node_indices = np.arange((nx + 1) * (ny + 1) * (nz + 1)).reshape(nx + 1, ny + 1, nz + 1)
    dof_tensor = np.zeros((nx + 1, ny + 1, nz + 1, 3), dtype=int)
    dof_tensor[..., 0], dof_tensor[..., 1], dof_tensor[..., 2] = node_indices * 3, node_indices * 3 + 1, node_indices * 3 + 2
    dof_tensor[nx, :, :, :] = dof_tensor[0, :, :, :]
    dof_tensor[:, ny, :, :] = dof_tensor[:, 0, :, :]

    n1, n2 = dof_tensor[:-1, :-1, :-1, :], dof_tensor[1:, :-1, :-1, :]
    n3, n4 = dof_tensor[1:, 1:, :-1, :], dof_tensor[:-1, 1:, :-1, :]
    n5, n6 = dof_tensor[:-1, :-1, 1:, :], dof_tensor[1:, :-1, 1:, :]
    n7, n8 = dof_tensor[1:, 1:, 1:, :], dof_tensor[:-1, 1:, 1:, :]
    edof_tensor = np.concatenate([n1, n2, n3, n4, n5, n6, n7, n8], axis=-1)

    active_mask = voxel > 0  # Filter out void regions
    edofMat = edof_tensor[active_mask]

    z_coords = np.linspace(dz / 2, thickness - dz / 2, nz) - thickness / 2.0
    z_grid = np.broadcast_to(z_coords[None, None, :], (nx, ny, nz))  # Z is the 3rd axis
    z_active = z_grid[active_mask]
    return edofMat, z_active, node_indices.size * 3

def homogenization_plate(voxel, E=2000.0, nu=0.3, thickness=10.0, Nx=1, Ny=1, Nz=1):
    nx, ny, nz = voxel.shape
    cell_size = thickness / Nz
    Lx = Nx * cell_size
    Ly = Ny * cell_size
    Lz = thickness
    dx, dy, dz = Lx / nx, Ly / ny, Lz / nz
    plate_area = Lx * Ly  # Normalization factor for LPS-H

    C = get_isotropic_elasticity(E, nu)
    Ke, Bs, detJ = compute_element_stiffness(C, dx, dy, dz)
    edofMat, z_active, total_dofs = build_tensor_dof_mapping(voxel, dx, dy, dz, thickness)

    iK, jK = np.repeat(edofMat, 24, axis=1).flatten(), np.tile(edofMat, (1, 24)).flatten()
    sK = np.tile(Ke.flatten(), edofMat.shape[0])
    K = coo_matrix((sK, (iK, jK)), shape=(total_dofs, total_dofs)).tocsr()  # Loop-free assembly of global stiffness

    E_macro = np.zeros((len(z_active), 6, 6))
    E_macro[:, 0, 0], E_macro[:, 1, 1], E_macro[:, 5, 2] = 1.0, 1.0, 1.0  # Apply unit membrane strains
    E_macro[:, 0, 3], E_macro[:, 1, 4], E_macro[:, 5, 5] = z_active, z_active, z_active  # Apply unit bending curvatures

    F_ele = sum([np.einsum('ji,kjl->kil', Bs[i], np.einsum('ij,kjl->kil', C, E_macro)) * detJ for i in range(8)])  # Local load via tensor contraction
    F = np.column_stack([np.bincount(edofMat.flatten(), weights=F_ele[:, :, c].flatten(), minlength=total_dofs) for c in range(6)])  # Global load assembly
    active_dofs = np.setdiff1d(np.unique(edofMat), np.unique(edofMat)[:3])  # Eliminate rigid body motions

    U = np.zeros((total_dofs, 6))
    K_active_gpu = cpsp.csr_matrix(K[active_dofs, :][:, active_dofs])
    F_active_gpu = cp.asarray(F[active_dofs, :])
    U_active_gpu = cp.zeros((len(active_dofs), 6))
    M_gpu = cpsp.diags(1.0 / K_active_gpu.diagonal())
    streams = [cp.cuda.Stream(non_blocking=True) for _ in range(6)]  # Initialize concurrent CUDA streams
    for c in range(6):
        with streams[c]:  # Launch asynchronous GPU solving for 6 load cases
            U_col_gpu, _ = cpspla.cg(K_active_gpu, F_active_gpu[:, c], M=M_gpu, tol=1e-6, maxiter=5000)  # Jacobi Preconditioner Conjugate gradient solver on GPU
            U_active_gpu[:, c] = U_col_gpu
    for stream in streams:
        stream.synchronize()  # Barrier synchronization
    U[active_dofs, :] = U_active_gpu.get()

    ABD = np.zeros((6, 6))
    for i_gp in range(8):
        Sigma = np.einsum('ij,kjl->kil', C, E_macro - np.einsum('ij,kjl->kil', Bs[i_gp], U[edofMat, :]))  # Recover true physical stress
        ABD[0:3, :] += np.sum(Sigma[:, [0, 1, 5], :], axis=0) * detJ / plate_area  # Extract A and B matrices (0th moment)
        ABD[3:6, :] += np.sum(Sigma[:, [0, 1, 5], :] * z_active[:, None, None], axis=0) * detJ / plate_area  # Extract D and B* matrices (1st moment)
    return (ABD + ABD.T) / 2.0  # Symmetrization for Maxwell-Betti reciprocity
\end{lstlisting}

\section{Comparison of LVS-H, LPS-H, and the Full-Scale Model}\label{sec:appendix_validation}
\begin{figure*}[htbp]
  \centering
  \includegraphics[width=\textwidth]{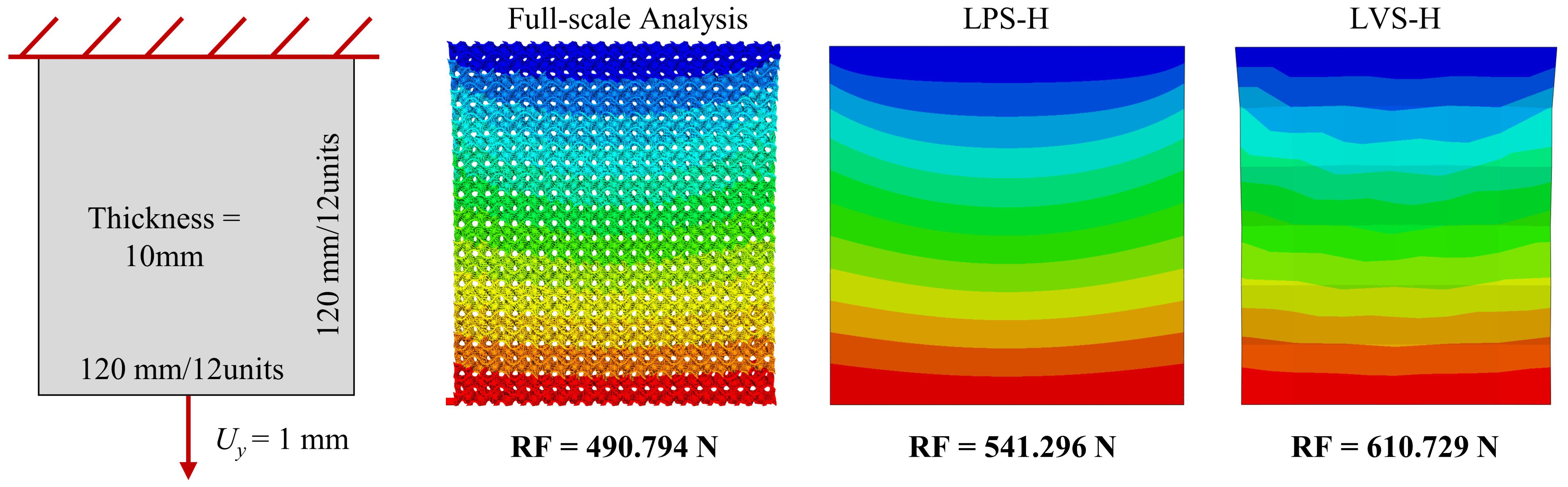}
  \caption{Comparison of the static responses predicted by the full-scale finite element model, LPS-H, and LVS-H. All three models use the same boundary conditions, with the upper boundary fixed and a $1$ mm displacement load applied to the lower boundary, and the support reaction is used to characterize the overall stiffness.}
  \label{fig:appendix_validation}
\end{figure*}

This appendix presents the static comparison results for a representative Lattice-skin Plate Structure with dimensions of $120 \times 120 \times 10$ mm under three modeling strategies: the full-scale finite element model, LPS-H, and LVS-H, as shown in \textbf{Figure \ref{fig:appendix_validation}}. To ensure fair comparison, the three models use the same boundary conditions and loading setup: the upper boundary along the $y$ axis is fully fixed, a displacement load of $1$ mm is applied to the lower boundary, and the overall support reaction of the structure is used as the effective stiffness indicator.

The calculated support reactions of the full-scale finite element model, LPS-H, and LVS-H are $490.794$ N, $541.296$ N, and $610.729$ N, respectively. Taking the full-scale finite element result as the reference, the relative deviation of LPS-H is $10.29\%$, whereas that of LVS-H reaches $24.43\%$. This shows that for finite-thickness Lattice-skin Plate Structures, LPS-H can represent the overall response of the real structure under displacement-controlled loading more accurately because it retains the boundary features of the free surfaces through the thickness. By contrast, LVS-H tends to overestimate structural stiffness because of the periodic continuation assumption introduced through the thickness, and therefore predicts a larger support reaction.

It should be noted that the full-scale finite element model is used only for benchmark validation, and its computational cost is much higher than that of the two homogenized models. The model contains more than $5 \times 10^6$ elements, requires a high-performance computing platform for solution, and has a peak memory demand of about $240$ GB. For Lattice-skin Plate Structures with complex internal topology and finite-thickness truncation effects, direct full-scale 3D simulation requires extensive preprocessing and meshing effort and also consumes substantial computational resources. In comparison, LPS-H retains the key thickness-boundary effects while greatly reducing the computational scale, making it a more efficient substitute model for engineering analysis and structural design.

\end{appendices}

\bmhead{Acknowledgements}
The computations in this paper were run on the Siyuan1 cluster supported by the Center for High Performance Computing at Shanghai Jiao Tong University.

\bmhead{Author contributions}
Zhongkai Ji and Dawei Li contributed to the conceptualization of the study. Zhongkai Ji contributed to the methodology, validation, and original draft preparation of the manuscript. Dawei Li contributed to funding acquisition, project administration, and manuscript review and editing. Yong Zhao contributed to the methodology, supervision, and manuscript review and editing. Wenhe Liao contributed to resources and supervision. Dawei Li (ldw@njust.edu.cn) and Yong Zhao (yongzhao@sjtu.edu.cn) are co-corresponding authors of this paper.

\bmhead{Funding}
This work was supported by the National Natural Science Foundation of China (Grant No. 52575292).

\bmhead{Data availability}
The data used in this study are available at \href{https://github.com/TopJournals/ThinWalledHomogenization}{https://github.com/TopJournals/ThinWalledHomogenization}.

\section*{Declarations}
\bmhead{Conflict of interest}
The authors declare that they have no known competing financial interests or personal relationships that could have appeared to influence the work reported in this paper.

\bmhead{Replication of results}
The code is available at \href{https://github.com/TopJournals/ThinWalledHomogenization}{https://github.com/TopJournals/ThinWalledHomogenization}.

\bibliography{total}% common bib file
\end{document}